\documentclass[11pt]{article}
\usepackage[usenames,dvipsnames]{color}
\usepackage{amsfonts}
\usepackage{amsmath}
\usepackage{amssymb}
\usepackage{graphicx}
\usepackage[pdftitle={DE clustering},pdfauthor={G Ballesteros},bookmarks=true,bookmarksnumbered=true]{hyperref}  
\usepackage{cite}

\usepackage{graphicx}
\usepackage{epsf}
\usepackage{graphics}
\usepackage{psfrag}
\usepackage{subfig}

\numberwithin{equation}{section}

\def\be{\begin{equation}}
\def\ee{\end{equation}}
\def\bea{\begin{align}}
\def\eea{\end{align}}

\newcommand{\rmd}{\mathrm{d}}

\newcommand{\eq}[1]{(\ref{#1})}

\def \ch{{\cal H}} 
\def \dc{\delta_m}
\def \dx{\delta_x}
\def \d{\delta}
\def \cshat{\hat{c_s}^2}

\def \cad{{c_a}^2}
\def \cs{{c_s}^2}
\def \fac{{\cal F}}
\def \dac{{\cal D}}
\def \vel{{\mathbf{u}}}

\def\ds{\displaystyle}

\long\def\symbolfootnote[#1]#2{\begingroup%
\def\thefootnote{\fnsymbol{footnote}}\footnote[#1]{#2}\endgroup}

\begin{document}

\setcounter{footnote}{0}
\setcounter{figure}{0}
\setcounter{table}{0}

\begin{center}

{\Large {\bf Non--linear dark energy clustering}}
\vspace{0.7cm}
\end{center}
\vspace{0.23cm}
\begin{center}
\symbolfootnote[0]{anselmi@pd.infn.it}\textbf{Stefano Anselmi}$^{(a)}$, \symbolfootnote[0]{ballesteros@pd.infn.it}\textbf{Guillermo Ballesteros}$^{(b,\, a,\, c)}$, \symbolfootnote[0]{pietroni@pd.infn.it}\textbf{Massimo Pietroni}$^{(c)}$\\ 
\vspace{0.25cm}
\small $a$. {\it \footnotesize  Dipartimento di Fisica ``G. Galilei'', Universit\`a degli Studi di Padova, via Marzolo
8, I-35131 Padua, Italy.}\\
\small $b$. {\it \footnotesize Museo Storico della Fisica e Centro Studi e Ricerche ``Enrico Fermi''.
Piazza del Viminale 1, I-00184 Rome. Italy}\\ \small $c$. {\it \footnotesize  INFN, Sezione di Padova, via Marzolo 8, I-35131 Padua, Italy} \\
\end{center}

\vspace{0.5cm}

\begin{abstract}

We consider a dark energy fluid with arbitrary sound speed and equation of state and discuss the
effect of its clustering on the cold dark matter distribution at the non--linear level. We write
the continuity, Euler and Poisson equations for the system in the Newtonian approximation.
Then, using the time renormalization group method to resum perturbative corrections at all orders, we
compute the total clustering power spectrum and matter power spectrum.
At the linear level, a sound speed of dark energy different from that of light modifies the power
spectrum on observationally interesting scales, such as those relevant for baryonic acoustic
oscillations. We show that the effect of varying the sound speed of dark energy on the non--linear corrections to the matter power spectrum is below the per cent level, and therefore these corrections can be well modelled by their counterpart in cosmological scenarios with smooth dark energy. We also show that the non--linear effects on the matter growth index can be as large as 10--15 per cent for small scales. 

\end{abstract}
\vspace{-0.2cm}

\newpage
\setcounter{page}{1}

\section{Introduction}
The accelerated expansion of the universe, first detected with Type Ia supernovae observations \cite{Riess:1998cb, Perlmutter:1998np}\,, is one of the major problems of cosmology. Today, there is ample evidence for it from the combination of supernovae data with the cosmic microwave background and baryon acoustic oscillations \cite{Komatsu:2010fb,Suzuki:2011hu}\,; and even from the cosmic microwave background alone \cite{Sherwin:2011gv}\,. The task of discriminating between different theoretical possibilities able to describe the acceleration is of the utmost importance not only for cosmology but also for the field of high energy physics. In this context, the ultimate question that we would like to answer is what is the dark energy behind the acceleration. 

It is well known that any smooth expansion history of the universe compatible with the observations can be described in the framework of general relativity by constructing an effective time dependent equation of state which is given by the time evolution of the Hubble parameter\symbolfootnote[2]{See for instance \cite{Bonvin:2006en} for an explicit expression.}. This means that a theory of modified gravity cannot be told apart from a model based on general relativity on the single basis of the determination of the background dynamics, even if done at very high precision. 

The expansion history also affects the way in which cold dark matter inhomogeneities grow to form structures in the universe. Clearly, a fast expansion slows the growth rate. The simplest equation to approach to this phenomenon in general relativity can be written, using conformal time, as:
\begin{align} \label{simplest}
\ddot\delta_m+\ch\dot\delta_m-\frac{3}{2}\Omega_m\ch^2\delta_m=0
\end{align}
The effect of the background on the matter perturbation $\delta_m$ is due to the time dependencies of the conformal Hubble parameter $\ch$ and the relative matter density $\Omega_m$\,. Therefore, the evolution of inhomogeneities provides an alternative tool to gain insight about the properties of dark energy. 

Since we know next to nothing about the actual physics causing the acceleration of the universe, we can adopt a phenomenological point of view and effectively describe it as a fluid in general relativity, as we do for matter. Such a fluid will also develop perturbations, as matter does. The time evolution of these perturbations will depend on the properties of the fluid: its equation of state, sound speed and anisotropic stress. The growth described by \eq{simplest} is then only valid in the case in which the dark energy fluid represents a cosmological constant. For any other case, the equation needs to be modified to take into account the effect of dark energy perturbations. These fluctuations change the right hand side of \eq{simplest}\,, modify the metric perturbations as well and obey another second order dynamical equation which also depends on the matter ones \cite{Ballesteros:2008qk}\,. It is unclear whether observations related to the perturbations can tell us if modified gravity or scalar field models within general relativity are to be preferred, but they are undoubtedly a great source of information to learn about the properties of the acceleration. 

Most theoretical studies related to dark energy fluctuations that have been done so far are in the framework of linear perturbation theory \cite{Dave:2002mn, Sandvik:2002jz, Bassett:2002fe, DeDeo:2003te, Doran:2003xq, Amendola:2003bz, Afshordi:2006ad, Kunz:2006wc, Creminelli:2008wc, Ballesteros:2008qk, Avelino:2008zz, Avelino:2008cu, Sapone:2009mb, Kunz:2009yx, Lim:2010yk, Koshelev:2010wz,Novosyadlyj:2010pg,Ansari:2011wv} and so are all the works on present constraints and forecasts for detection \cite{Bean:2003fb, Weller:2003hw, Hu:2004yd, Hannestad:2005ak, Corasaniti:2005pq, Kunz:2006ca, Takada:2006xs, Amendola:2007rr, Mota:2007sz, TorresRodriguez:2007mk, Xia:2007km, dePutter:2010vy, Ballesteros:2010ks, Sapone:2010uy, Li:2010ac, Ayaita:2011gp}\,,  but given the increasing precision of present and future planned probes, the relevance of non--linear effects has to be taken into account. In particular, the power spectrum of matter fluctuations is sensibly affected by non--linearities for wavenumbers larger than $0.1 h\, \text{Mpc}^{-1}$\,, where a good deal of information can be gained from galaxy surveys. In this work, we asses the importance of these effects for the growth of matter fluctuations in the presence of a dark energy fluid. We focus on the effect of the sound speed, assuming that the dark energy has no anisotropic stress. 

Presently, there are no observational constraints on the sound speed of dark energy because the available data are not powerful enough. However, it has been shown that combining the cosmic microwave background with large scale structure data it will be possible to set a lower bound on it in a near future \cite{Hu:2004yd,Ballesteros:2010ks,Sapone:2010uy,Ayaita:2011gp}\,. The precise bounds will be affected by the effect of the non--linear evolution of the perturbations.

In \cite{Bjaelde:2010qp} a ``quasi--nonlinear'' approach was considered to study the spherical collapse approximation with dark energy fluctuations with a small sound speed. Maximal clustering of the dark energy perturbations occurs if their sound speed is zero. The spherical collapse was analysed for such a case in \cite{Creminelli:2009mu}\,. Second order solutions for the energy density and velocity perturbations in Eulerian perturbation theory were found, also for a vanishing sound speed, in \cite{Sefusatti:2011cm}\,. The lowest order (tree--level) contributions to the bispectrum were also calculated in the same work. 

Dark energy fluctuations are typically suppressed with respect to those of matter and this makes their detection difficult. The change that they produce on the matter growth $\delta_m/a$ with respect to what is expected in a smooth  $w\text{CDM}$ cosmology without dark energy perturbations can be estimated from the growth index $\gamma$\, \cite{Linder:2005in}\,, which is modified by $\sim 5\%$ at most \cite{Ballesteros:2008qk}\,. The matter growth itself, normalized at $z\rightarrow\infty$ and evaluated today, changes by at most $1\%$\,, depending on the scale. These small changes are however large enough to be comparable to the predicted sensitivities and hence one could hope to connect with dark energy perturbations by measuring the matter growth. These results are based on linear perturbation theory and therefore it makes sense to see how they can be modified when non--linear corrections are taken into account. We will see that indeed the non--linear effects greatly affect (to the level of 10--15 $\%$) the matter growth index for small scales and therefore they should be accounted for when the index is used (for model comparison, for instance).

We concentrate on scales below the Hubble horizon, concretely on the relevant ones for the observation of the matter power spectrum with large scale structure surveys, and use the time renormalization group (TRG) \cite{Pietroni:2008jx}\,, which is a semi--analytical method to solve the Euler, continuity and Poisson equations resumming classes of perturbative corrections at all orders. In this work, we set a useful theoretical framework to deal with non--linearities in the Newtonian approximation for a mixture of matter and a non--pressureless fluid characterized by a non--adiabatic sound speed. We write the non--linear continuity and Euler equations for subhorizon scales and build a consistent approximation of the non--linear terms that allows the application of the TRG. We also explain the relevance of the sound speed and where it comes into play. While in the case of pure pressureless quintessence, the sound horizon of dark energy has zero length and fluctuations cluster for all scales smaller than $\ch^{-1}$\,; if the sound speed is different from zero, scale dependent effects become relevant already at the linear level. When the sound speed is non--zero, at this level, the time variation of the pressure of dark energy becomes important and, in consequence, the velocity perturbations of dark matter and dark energy are not equal (the two fluids are not comoving), in contrast to the case of pressureless quintessence analysed in \cite{Sefusatti:2011cm}\,. 

We apply the TRG to obtain the non--linear power spectrum of density fluctuations. Since at the linear level dark energy perturbations are suppressed with respect to the matter ones by a $1+w$ factor, a non--linear treatment is strictly needed only for matter perturbations. This observation simplifies the implementation of the TRG approach considerably and is at the basis of the approximation that will be described in Section \ref{appsol}\,. Our results indicate that the hierarchy between the two components that occurs at the linear level is maintained at the non--linear one, confirming the consistency of the approximation.

As already pointed out in \cite{Creminelli:2009mu, Sefusatti:2011cm}\,, if the dark energy component clusters on observable scales, it may become problematic to disentangle its effects from those of cold dark matter, e.g. in lensing measurements, so that a case by case discussion of the different observables is needed. Therefore, it makes sense to consider also the power spectrum of the full clustering fluid, i.e. cold dark matter plus dark energy. We comment on this issue examining the growth function for the matter and full spectra at the linear and non--linear levels.

The structure of the paper is as follows. In section \ref{new} we describe the Newtonian approximation for a fluid with pressure, writing the Euler, continuity and Poisson equations and checking their consistency with linear perturbation theory in general relativity. We show that no pressure terms appear in the Poisson equation and explain the importance of time derivatives that have been neglected in earlier works. In Section \ref{nlp} we present the non--linear perturbation equations, which we provide to the level that we will need for our numerical computations. Section \ref{dscales} is devoted to explain in more depth the general approximations that we have adopted. We review the TRG method and particularize it for our case of interest in Section \ref{trg}. The next section, \ref{appsol}\,,  describes the approximation that we use to reduce the number of equations in the TRG for our numerical solution, which is given in Section \ref{numerical}\,. Section \ref{conclusions} corresponds to the conclusions. The Appendix \ref{sv} discusses the formulas of \cite{Sefusatti:2011cm} in the context of the equations of Section \ref{new}. In our work we also use these formulas in Section \ref{numerical} for the purpose of comparison.

\section{The Newtonian approximation} \label{new}
We consider a system composed of two cosmological fluids. One of them is pressureless and represents baryons and dark matter together. The other fluid corresponds to the dark energy component of the universe. Since we are interested in studying the non--linear effects of dark energy on the cold dark matter power spectrum and we focus on the epochs of matter and dark energy domination, we do not need to include other components such as radiation. We will label quantities relative to the matter component with a subindex `$_m$' and those referring to dark energy will be tagged with an `$_x$'\,. Whenever that an equation involving fluid quantities without labels appears in the text, it will apply to any of the two components.

In this section we will present the equations that describe the system in the Newtonian approximation. We will specify the properties of the fluids and describe our approximations and their range of validity, which will be further explained in Section \ref{dscales}\,. Although in the following we keep the non--linearities in the dark energy perturbations, in our numerical calculations, described in Section \ref{appsol}\,, we will treat dark energy perturbations linearly, an approximation motivated by equation \eq{supsound} below.

\subsection{Cosmological fluid dynamics with pressure}
 Our starting point is the following couple of equations:
\begin{align} \label{corrcont}
\left(\rho+\vel^2 P\right)^{\displaystyle \cdot}+3\ch(\rho+P)+\nabla\cdot\left[\left(\rho+P\right)\vel\right]&=0\\
\dot\vel+\ch\vel+\left(\vel\cdot\nabla\right)\vel+\frac{\nabla P+\vel\dot P}{\rho+P}+\nabla\phi_N&=0 \label{correuler}
\end{align}
The density and pressure of the fluid are denoted by $\rho$ and $P$ respectively and the peculiar velocity is represented by $\vel$\,. Overdots denote derivatives with respect to conformal time, $\ch$ is the conformal Hubble factor and $\nabla$ is the three--dimensional gradient for comoving spatial coordinates.
These two equations are meant to improve the classical continuity and Euler equations \cite{Peebles:1980book} in their application to describe a fluid with pressure in an expanding cosmological background. The continuity equation \eq{corrcont} differs from the classical version of \cite{Peebles:1980book} because it includes the pressure inside the divergence term and in the time derivative of the first term. Besides, the Euler equation \eq{correuler} also contains the time derivative of the pressure, which was not originally present either \cite{Peebles:1980book}\,. The quantity $\phi_N$ represents the Newtonian potential and encodes the effect of the metric perturbations.

The continuity and Euler equations describe local energy and momentum conservation. They can be obtained from general relativity applying the covariant conservation equations \cite{Wald:1984rg}
\be \label{cov}
T^{\mu\nu}_{\quad;\,\mu}=0
\ee
to an energy--momentum tensor of the form 
\be \label{tmunu}
T^{\mu \nu}=\left(\rho+P\right)u^\mu u^\nu+P\, g^{\mu \nu}\,,
\ee
assuming that metric perturbations are small (and hence $\phi_N\ll 1$) and that the peculiar velocity is much smaller than the speed of light: $|\vel|\ll 1$\,. Concretely, these equations can be obtained writing the four--velocity of the fluid as $u^\mu=(1,\vel)$ in \eq{tmunu} (since the Lorentz factor $\gamma\simeq 1$) and neglecting high order metric corrections. These lines were followed in \cite{Sefusatti:2011cm} neglecting the time derivative of $\vel^2 P$ in the  continuity equation\symbolfootnote[2]{This term was also absent in \cite{Creminelli:2009mu} and \cite{Bjaelde:2010qp}\,.} . However, in order to do a fully consistent treatment of the perturbations, this term has to be included. 

If we expand this derivative using $\eq{correuler}$ to replace $\dot\vel$ in \eq{corrcont} and we keep the terms that contain up to two perturbations, we obtain:
\be \label{continuity}
\dot\rho+3\ch(\rho+P)+\nabla\cdot\left(\rho\,\vel\right)+P\,\nabla\cdot \vel+\frac{\rho-P}{\rho+P}\,\vel\cdot(\nabla P+\vel\dot P) -2P\,\vel\cdot\left(\ch\vel+\nabla\phi_N\right)=0
\ee
This equation shows explicitly that there are terms of second order in the perturbations that cannot be neglected by throwing away the $\vel^2 P$ term in \eq{corrcont}\,. Notice also that, even if $|\vel|\ll 1$\,, it is not possible to neglect the term $\vel\dot P$ in \eq{correuler} with the argument that for small velocities $\nabla P\gg\vel \dot P$\,. This is because $\vel \dot P/(\rho+P)$ contains (after applying the continuity equation for the background) a term $-3w\ch\vel$ that is of the same order of magnitude as $\ch\vel$\,, which is the second term of \eq{correuler}\,. 

If we had also included the effect of the Lorentz factor $\gamma$ we would have got extra non--linear corrections proportional to $\vel\cdot\dot\vel$\,, which we neglect. These corrections can be neglected in our treatment because dark energy perturbations are suppressed with respect to those of matter (as we will later explain in more detail). Otherwise, these corrections should be taken into consideration.

We will split the total density and pressure according to 
\begin{align} \label{fullrho}
\rho &= \bar\rho\left(1+\delta\right)\\ \label{fullp}
P &=\bar P\left(1+\Pi\right)\,,
\end{align}
where the quantities with an overbar refer to the  background (time dependent) homogeneous cosmology and we define the perturbations
\begin{align} 
\delta\rho &= \bar \rho\, \delta \\
\delta P &= \bar P\, \Pi\,,
\end{align}
which in general are functions of time and space coordinates. The Newtonian potential $\phi_N$ can be expressed in terms of the total density and velocity perturbations through the Poisson equation
\be \label{poisson}
\nabla^2\phi_N=4\pi G a^2\sum_{\alpha\, =\, m,\, x}\left(\delta\rho_\alpha-3\left(\bar\rho_\alpha+\bar P_\alpha\right)\ch\chi_\alpha\right)
\ee
Here, $G$ is Newton's gravitational constant, $a$ represents the scale factor of the universe, the index $\alpha$ sums over the contributions of matter and dark energy and we denote by $\chi$ the velocity potential of a fluid:
\be
\vel=\nabla\chi
\ee
Therefore, we are assuming that both components, matter and dark energy, are irrotational. This assumption is well justified for matter because the curl of its peculiar velocity decays as $a^{-1}$ at the linear level. For a fluid with pressure, if we define $\mathbf{w}=\nabla\times \vel$\,, we have $\dot{\mathbf{w}}=(-1+3w)\ch \mathbf{w}$, which implies that the decay is even faster for negative $w$, also at the linear level. Then, for simplicity, we will assume that we have zero dark energy vorticity at all times.

It is worth noticing that the Poisson equation \eq{poisson} does not contain any pressure terms, in contrast with what was written before in other related works  \cite{Creminelli:2009mu,Bjaelde:2010qp,Sefusatti:2011cm} and also much earlier in \cite{Peebles:1980book}\,. What remains of this section is devoted to check the consistency of the equations with linear perturbation theory in general relativity, using the synchronous and conformal Newtonian gauges for illustration. In Subsection \ref{eqcs} we will review the concept of non--adiabatic sound speed and emphasize the need for using its rest frame value. 

\subsection{The Poisson equation}

Let us start with the Poisson equation \eq{poisson}\,. In the conformal Newtonian gauge, within the context of linear perturbation theory in general relativity, the Einstein equations in Fourier space (see \cite{Ma:1995ey}, for instance) imply that the two metric scalar degrees of freedom are equal to each other and given by: 
\be \label{psn}
k^2\phi=-4\pi G a^2\sum_\alpha\left(\delta\rho_\alpha^{(c)}+3\left(\bar\rho_\alpha+\bar P_\alpha\right)\frac{\ch^2}{k^2}\frac{\theta_\alpha^{(c)}}{\ch}\right)
\ee
for any system of fluids with zero anisotropic stress. Here, the labels `$^{(c)}$' refer to quantities in the conformal Newtonian gauge, $k$ is the wavenumber  and $\theta$ denotes the velocity divergence, defined as $(\bar{\rho}+\bar{P})\theta \equiv i k^j \delta T^0{}_{\!j}$\,, and which we can link to $\vel$ in the Newtonian approximation by
\be 
\theta=\nabla\cdot\vel
\ee
It is straightforward to obtain \eq{psn} from the perturbed Einstein equations
\begin{align}
\delta G_{\mu\nu}=8\pi G\delta T_{\mu\nu}
\end{align}
Following the notation of \cite{Ma:1995ey}\,, we write the FRW metric for scalar perturbations in the Conformal Newtonian gauge as
\begin{align}
\label{conformal}
    ds^2 = a^2(\tau)\left\{ -(1+2\psi)d\tau^2 +
        (1-2\phi)dx^i dx_i \right\}
\end{align}
If we express the traceless component of $T^i{}_{\!j}$ as $
\Sigma^i{}_{\!j} \equiv T^i{}_{\!j}-\frac{1}{3}\delta^i{}_{\!j} T^k{}_{\!k}
$
and define the anisotropic stress
$
(\bar{\rho}+\bar{P})\sigma \equiv -(\hat{k}_i\hat{k}_j
	-  \frac{1}{3} \delta_{ij})\Sigma^i{}_{\!j}\,
$, where the hats stand for unit length vectors,
the Einstein equations read \cite{Ma:1995ey}\,:
\begin{align}
    k^2\phi + 3\ch \left( \dot{\phi} + \ch\psi
	\right) &= -4\pi G a^2 \delta \rho^{(c)}
	\label{eincona}\\
    k^2 \left( \dot{\phi} + \ch\psi \right)
	 &= 4\pi G a^2 (\bar{\rho}+\bar{P}) \theta^{(c)}
	 \label{einconb}\\
    \ddot{\phi} + \ch (\dot{\psi}+2\dot{\phi})
	+\left(2\dot\ch + \ch^2\right)\psi
	+ \frac{k^2}{3} (\phi-\psi)
	&= 4\pi G a^2 \delta P^{(c)}
	\label{einconc}\\
    k^2(\phi-\psi) &= 12\pi G a^2 (\bar{\rho}+\bar{P})\sigma\,,
	\label{eincond}
\end{align}
where the terms on the right hand side have to be understood as the sums for all the fluid components (dark matter and dark energy in our case).
Combining \eq{eincona} and \eq{einconb} one obtains precisely \eq{psn}\,. Then, under the assumption that $\sigma=0$ (no anisotropic stress) we have $\phi=\psi$\,. Therefore, in order to have a pressure component in the Poisson equation, the fluid would have to have non-zero anisotropic stress (which is a gauge invariant quantity). Notice also that during matter domination $\ch=2/\tau$ and therefore $2\dot\ch+\ch^2$ is zero. This clearly indicates that the pressure perturbation in \eq{einconc} is of the order of the terms involving the time derivatives of the potential, which in the Newtonian approximation can be neglected.

Using the Friedmann equation for $\ch$ we can express \eq{psn} as follows:
\begin{align} \label{psnf}
-k^2\phi=\frac{3}{2}\ch^2\sum_\alpha\Omega_\alpha\left(\delta_\alpha^{(c)}+3\left(1+w_\alpha\right)\frac{\ch^2}{k^2}\frac{\theta_\alpha^{(c)}}{\ch}\right)
\end{align}
The equation \eq{poisson} is the counterpart of \eq{psnf} in real space. We will later argue that the only relevant term of the Poisson equation in the Newtonian approximation is the one that depends on the density perturbations $\delta_\alpha\,$. But before that, we are going to introduce the parameters that characterize the properties of the dark energy fluid.
\subsection{Equation of state and sound speed} \label{eqcs}
We will now introduce the parameters that relate the pressure and the energy density in the fluids we consider. We assume that the background pressure and density of each fluid are linked through
\be
\bar P=w \bar \rho
\ee
and, similarly, we write the following relation for their perturbations 
\be 
\delta P =\cs\, \delta \rho \label{soundspeed}
\ee
These two expressions define the equation of state parameter $w$ and the sound speed $c_s$ for each fluid. By construction, the equation of state depends only on time, while the speed of sound may in principle also vary in space. If we imposed that the sound speed is solely a function of time, the fluid would be instantaneously barotropic at the background and perturbation levels.
Notice that the relative density and pressure perturbations are related through
\be 
\cs\, \delta = w\, \Pi
\ee
Our matter fluid will be assumed to have exactly zero $w$ and sound speed, while for dark energy we will allow these quantities to take constant values and assume $w<-1/3$\,. For our numerical treatment of the non--linear effects, the assumption that $w$ and $\cshat$ are constant is the simplest choice one can make in absence of much information about the nature of dark energy and it is a phenomenologically reasonable one since in many field models both quantities are very slowly varying during the last stages of the evolution of the universe. Lacking a specific preferred model of dark energy, working with constant parameters is a simple and justified choice.  

We restrict our analysis to fluids that interact only gravitationally and therefore satisfy a conservation equation of the form
\be \label{cons}
\dot{\bar \rho}+3(1+w)\ch\bar\rho=0\,,
\ee
which is nothing else than the background (linear) part of \eq{corrcont}\,.
In consequence,
\be \label{adiab}
\dot w=3(1+w)\ch\left(w-{c_a}^2\right)\,,
\ee
where we define the adiabatic sound speed of the fluid to be given by
\be 
\dot{\bar P}={c_a}^2\dot{\bar \rho}
\ee
Clearly, ${c_a}^2=w$ if the equation of state parameter is constant. In the general case that $c_s \neq c_a$\,, it is said that the fluid is not adiabatic. For an interpretation of the non--adiabatic sound speed $c_s$ in terms of entropy perturbations we refer the reader to \cite{Bean:2003fb}\,.

In the framework of linear perturbation theory in general relativity, the sound speed defined in \eq{soundspeed} is a gauge dependent quantity. Since we are interested in studying its effect in the clustering of matter, it is convenient to replace $c_s$ by a gauge independent quantity that can be potentially measured. This can be achieved by using the following relation in Fourier space \cite{Bean:2003fb}:
\be \label{rf}
\left(\cshat-c_{s}^2\right)\delta=3(1+w)\left(\cad-\cshat\right)\frac{\ch}{k^2}\theta\,,
\ee
which (in this form) is valid in the conformal Newtonian and synchronous gauges at linear level and gives us the rest frame (gauge independent) sound speed $\cshat$\,. In order to do a fully consistent second order treatment of dark energy perturbations we would have to modify the equation \eq{rf} to include second order terms. However, since we will treat these perturbations linearly in our numerical analysis (and all the non--linearities will be in the matter component), the equation \eq{rf} will be enough for our purposes. 

Let us finally remark that in the case of negative $w$\,, the sound speed of the fluid has to be non--adiabatic in order to avoid pathological instabilities. For a detailed study on this issue see \cite{Creminelli:2008wc}\,. In this work we will assume $0\leq \cshat\leq 1$\,. As mentioned in the Introduction, using cosmic microwave background and large scale structure data, it will be possible to set a lower bound on the sound speed of dark energy in a near future \cite{Hu:2004yd,Ballesteros:2010ks,Sapone:2010uy,Ayaita:2011gp}\,.

\subsection{Linear perturbations and consistency with general relativity} \label{matching}
As we have mentioned earlier, at the level of linear perturbations, the results obtained from the Newtonian approximation should match those of general relativity for scales smaller than $\ch^{-1}$\,. Before solving our full Newtonian equations \eq{continuity} and \eq{correuler}\,, we should check that this is indeed the case. To do the matching between the Newtonian approximation and general relativity we work in Fourier space and use $\cshat$ inside the Euler and continuity equations. Expressing the energy density and the pressure in \eq{corrcont} by \eq{fullrho} and \eq{fullp}\,, and then making use of \eq{rf} to introduce the rest frame sound speed, we get that the linear parts of these equations for the case of constant $w$ are
\begin{align} \label{linearcontnew}
\dot\delta &= -(1+w)\theta-3\left(\cshat-w\right)\ch\delta-9(1+w)\left(\cshat-w\right)\frac{\ch^2}{k^2}\theta\\ \label{lineareulernew}
\dot\theta &=-\left(1-3\cshat\right)\ch\theta +\frac{\cshat
k^2}{1+w} \delta+k^2\phi_N
\end{align}
These equations look very similar to the ones of general relativity for linear perturbations in the conformal Newtonian gauge:
\begin{align} \label{linearcont}
\dot\delta^{(c)} &= -(1+w)\left(\theta^{(c)}-3\dot\phi\right)-3\left(\cshat-w\right)\ch\delta^{(c)}-9(1+w)\left(\cshat-w\right)\frac{\ch^2}{k^2}\theta^{(c)}\\
\dot\theta^{(c)} &=-\left(1-3\cshat\right)\ch\theta^{(c)} +\frac{\cshat
k^2}{1+w} \delta^{(c)}+k^2\phi \label{lineareuler}
\end{align}
Already from here, we can immediately guess that in order to get the Newtonian approximation we need to identify $\phi$ and $\phi_N$ and be able to neglect the contribution of $\dot\phi$ in \eq{linearcont}\,. Indeed, using \eq{psnf} we see that the contribution of $\dot\phi$ in \eq{linearcont} gives contributions in $\delta$ and $\theta$ that are suppressed by powers of $\ch^2/k^2$ and $\ch^4/k^4$\,, respectively.

It was found in \cite{Ballesteros:2010ks} that the relative energy densities of matter and dark energy in the regime of pure matter domination ($\Omega_m=1$) are related by 
\be \label{supsound}
\dx=(1+w)\frac{1-2\cshat}{1-3w+\cshat}\,\delta_m\,,\quad \ch_s\gg k\gg\ch
\ee
for scales smaller than $\ch^{-1}$ (where the density perturbations are approximately gauge invariant\symbolfootnote[2]{See \cite{Ma:1995ey} or \cite{Ballesteros:2010ks}, for example.}) and larger than the sound horizon $\ch_s^{-1}=\hat{c}_s\ch^{-1}$\,. This expression can be obtained using the synchronous gauge by looking for the possible solutions of the second order differential equation for $\delta_x$ and taking into account that the growing mode for cold matter perturbations evolves in time as $\delta_m\propto k^2\tau^2$ during matter domination. The matter perturbations act as a source term for dark energy fluctuations and it is this source term that originates the solution \eq{supsound}\,.

Since the Newtonian approximation applies in this range of scales, the solution \eq{supsound} and the (gauge invariant) density perturbations computed in general  relativity (in any gauge) should be recovered from \eq{linearcontnew} and \eq{lineareulernew} in combination with \eq{psnf}\,. A good consistency test to check that this is indeed the case can be done from the second order differential equation for $\delta$. To obtain this equation, first in the conformal Newtonian gauge, we differentiate the continuity equation \eq{linearcont} with respect to conformal time and use the Euler equation \eq{lineareuler} and the continuity equation itself in the result. In this way we can eliminate the velocity divergence $\theta^{(c)}$\,. What we get is an equation (valid at any scale) that only contains $\delta^{(c)}$ and its first and second derivatives on the left hand side; and  the metric potential $\phi$ and its derivatives on the right hand one:
\begin{align}  \label{deper} \nonumber
\ddot\d^{(c)}&+\left[3\left(\cshat-w\right)\ch-\fac\right]\dot\d^{(c)}
-\frac{3}{2}\left(\cshat-w\right)\left[\left(1+3w\Omega_x-6\cshat\right)\ch^2
  +2 \ch\fac\right]\d^{(c)} +\cshat k^2
\d^{(c)}\\&=(1+w)\left(3\ddot\phi-3\fac\dot\phi-\dac\phi\right) , 
\end{align}
where 
\begin{align}
\dac&=k^2+9(\cshat-w)\ch^2 \\ \label{fac}\fac &=
-9\left(1+3w\Omega_x\right)\frac{\cshat-w}{\dac}\ch^3-(1-3\cshat)\ch
\end{align}
Notice that if we follow the same procedure using the equations \eq{linearcontnew} and \eq{lineareulernew} instead, we arrive to an expression which has formally the same functional structure as \eq{deper} with the difference that the derivatives of $\phi_N$ are absent. This is consistent with our previous guess that the time derivatives of $\phi$ must be negligible in the Newtonian limit. For scales $k\gg\ch$\,, the right hand side of \eq{deper} becomes equal to 
\be \label{deperight}
(1+w)\left(3\ddot\phi+3\left(1-3\cshat\right)\ch\dot\phi-k^2\phi\right)=(1+w)\left(\ddot\dc^{(c)}+\ch\dot\dc^{(c)}-9\cshat\ch\dot\phi\right)\,,\quad k\gg\ch
\ee
If we had done analogous operations working in the synchronous gauge, we would have obtained the same functional form for the left hand side of \eq{deper} (as explained in \cite{Ballesteros:2008qk, Ballesteros:2010ks}) and the following sum
\be 
(1+w)\left(\ddot\dc^{(s)}+(1-3\cshat)\ch\dot\dc^{(s)}\right)\,,\quad k\gg\ch
\ee
at the right hand side, where $\dc^{(s)}$ stands for the matter density perturbation in the synchronous gauge. Taking into account that the gauge differences of dark energy and dark matter perturbations are related by
\be 
\delta_x^{(c)}-\delta_x^{(s)}=(1+w)\left(\delta_m^{(c)}-\delta_m^{(s)}\right)
\ee
and expressing the time derivative of the potential as
\be \label{phid}
\dot\phi=\frac{k^2}{9\ch}\left(\delta_m^{(s)}-\delta_m^{(c)}\right)+\frac{1}{3}\dot\delta_m^{(c)}
\ee
one can check that the second order differential equation in the synchronous gauge \cite{Ballesteros:2010ks} is equal to \eq{deper} in the limit $k\ll\ch$ (up to negligible terms that are suppressed by factors of $\ch^2/k^2$ or higher powers)\,. Therefore, both gauges do indeed give the same result for subhorizon scales, as expected. In addition, notice that for subhorizon scales and in matter domination $\dot\delta_m=\delta_m\ch$ and therefore \eq{phid} tells us that  $\dot\phi\rightarrow 0$ and \eq{supsound} is also valid in the Newtonian approximation. In consequence, we see that the conformal Newtonian gauge is the best suited to arrive to the equations in the Newtonian approximation and we have checked the consistency of these with the synchronous gauge at small scales. From the arguments above, we can expect that working with the following Poisson equation should be an excellent approximation:
\be \label{psn2}
-k^2\phi_N=\frac{3}{2}\ch^2\sum_{\alpha\, =\, m,\, x}\Omega_\alpha\delta_\alpha\,,
\ee
because the contributions of the velocity divergences $\theta_\alpha\sim\ch\delta_\alpha$ from \eq{psnf} to the Euler equation  is suppressed (see also Section \ref{dscales})\,.

Notice also that looking at \eq{linearcont} or \eq{linearcontnew} one could naively expect that it would be correct to neglect the term in $\theta$ that is proportional to $\ch^2/k^2$ in those equations. However, such an approximation is not consistent for early times because it fails to reproduce the result \eq{supsound}\,, as one can easily check. The subtle reason behind this fact is the presence of the term proportional to $\cshat k^2\delta$ in the Euler equation. This feature is characteristic of fluids with pressure and therefore does not occur for pure dust. Moreover, let us remark that it is this same term in \eq{deper} that allows the equations to be consistent in different gauges. 

We can also express the velocity divergences in terms of the matter perturbation deep in the epoch of matter domination (for $\Omega_m$=1)\,. One can check, using \eq{linearcontnew}\,, that in the Newtonian approximation the relation is
\begin{align} \label{velrel}
\theta_x=\left(-1+\frac{6\,\cshat\left(\cshat-w\right)}{1-3w+\cshat}\right)\ch\delta_m\,,\quad \ch_s\gg k\gg\ch\,.
\end{align}
Clearly, the result obtained from \eq{linearcont} neglecting $\dot\phi$ (which we can do for those scales) is the same. This relation, together with \eq{supsound}\,, will serve us to define the initial conditions in our numerical analysis.

Having seen that our equations are consistent with general relativity not only at the background level \eq{cons} but also for linear perturbations around it (taking the limit from two different gauges), we can move on to study the non--linear dynamics.

\section{Non--linear perturbations} \label{nlp}
To find the non--linear continuity and Euler equations for the perturbations, it is convenient to write 
\begin{align}
\nabla P &={\bar\rho}\,\nabla\left(\cs\,\delta\right)\\
\dot P &=\left(\bar\rho\,\cs\,\delta\right)^\cdot+\cad\dot{\bar\rho}
\end{align}
and then use \eq{rf} together with the Einstein equations for zero intrinsic curvature 
\begin{align}
3\ch^2 &=8\pi G\,a^2\,\sum_\alpha\bar\rho_\alpha\\
3\dot\ch &=-4\pi G\,a^2\,\sum_\alpha\left(\bar\rho_\alpha+3\bar P_\alpha\right)
\end{align}
In Fourier space, the result is:
\begin{align}
\nonumber \label{eqc}
\dot\delta(\mathbf{k}) &+ 3\left(\cshat-w\right)\ch\delta(\mathbf{k})+(1+w)\left(1-9\left(w-\cshat\right)\frac{\ch^2}{k^2}\right)\theta(\mathbf{k})
\\&+\left(1+\cshat\right)\int  d^3\mathbf{p}\,d^3\mathbf{q}\,\delta_D(\mathbf{k}-\mathbf{p}-\mathbf{q})\alpha(\mathbf{q},\mathbf{p})\theta({q})\delta({p})+\mathcal{O}(2)=0\\\nonumber \label{eqe}
\dot\theta(\mathbf{k})&+\left(1 -3\cshat\right)\ch\theta(\mathbf{k})-\frac{\cshat k^2}{(1+w)}\delta(\mathbf{k})-k^2\phi_N
\\ &+(1-\cshat)\int d^3\mathbf{p}\,d^3\mathbf{q}\,\delta_D(\mathbf{k}-\mathbf{p}-\mathbf{q})\beta(\mathbf{q},\mathbf{p})\theta({q})\theta({p})+\mathcal{O}(2)=0\,,
\end{align}
where $\mathcal{O}(2)$ stands for the second order terms that we have not written explicitly. In order to compute all the contributions to those terms we would need to obtain the completion of \eq{rf} at second order, although some contributions can be readily obtained using just \eq{rf}. Our aim in this work is to compute the non--linear corrections to the power spectra (of matter and matter plus dark energy) so in principle, all the $\mathcal{O}(2)$ terms should be included. However, given that dark energy perturbations are suppressed with respect to those of matter, we will not need the second order terms for dark energy in our numerical calculations and the expressions above will be enough. This is justified by the expression \eq{supsound}\,, which indicates that for the values of $w$ and $\cshat$ that we are interested in, $\delta_x$ is roughly an order of magnitude smaller than $\delta_m$\,. The functions of momenta,  $\alpha$ and $\beta$\,, that appear in the explicit integrals of \eq{eqc} and \eq{eqe} are the usual ones for dust (see for instance \cite{Bernardeau:2001qr}):
\be
\alpha(\mathbf{q},\mathbf{p})=\frac{(\mathbf{q}+\mathbf{p})\cdot \mathbf{q}}{q^2}\,,\quad\quad \beta(\mathbf{q},\mathbf{p})=\frac{(\mathbf{p}+\mathbf{q})^2(\mathbf{q}\cdot\mathbf{p})}{2q^2\,p^2}\,.
\ee
The equations \eq{eqc} and \eq{eqe} become the standard ones for matter taking $w$ and $\cshat$ to be equal to zero. This is so because the second order corrections $\mathcal{O}(2)$ are either proportional to the sound speed $\cshat$ or the equation of state $w$\,.
\section{Distance scales and non--linear power counting} \label{dscales}
In this section we explain in more depth our approximations and comment on the relevance that the sound horizon of dark energy has on them.

We are interested in scales much smaller than the Hubble distance, i.e. $k\gg\ch$\,, which is the region of validity of the Newtonian approximation. However, $\ch^{-1}$ is not the only relevant scale in the problem; apart from it we have to consider the sound horizon of dark energy, which is defined as $\ch_s^{-1}=\hat{c}_s\ch^{-1}$\,. If the rest frame sound speed is equal to the speed of light, the sound horizon and the Hubble scale coincide. This is what happens, for instance, for quintessence models with a canonical kinetic term. However, in the general situation in which $\ch_s\neq\ch$ there will be two regions of interest for us, characterized by wave numbers $k$ larger o smaller than $\ch_s$\,. At the linear level, for scales smaller than the sound horizon ($k>\ch_s$)\,, dark energy perturbations display oscillations that do not occur outside it.

The fact that we will focus on $k\gg\ch$ at all times, allows us to make a power counting scheme to guess the relevance of the different terms that appear in the continuity, Euler and Poisson equations. It is well known that for scales smaller than the Hubble distance\,, $\theta\sim\ch\delta$  at the level of linear perturbations. This is the classical linear growth behaviour for matter perturbations, but it is also true for dark energy fluctuations, as can be checked neglecting the non--linear terms. 

If the relative energy densities $\delta_\alpha$ remain smaller than 1, the non--linear terms will generically give small contributions. Let us first look at the continuity equation. Going to Fourier space and using the Poisson equation, the value of $\dot\delta/\ch$ is determined by linear contributions of order $\Delta_{c(l)}$ and non--linear ones of order $\Delta_{c(nl)}$ where  
\be 
\Delta_{c(l)}\sim \mathcal{O}\left(\delta\right)\times\left(\mathcal{O}\left(1\right)+\mathcal{O}\left(\ch^2/k^2\right)\right)
\ee
and 
\be 
\Delta_{c(nl)}\sim\mathcal{O}\left(\delta\right)\times\left(\Delta_{c(l)}+\mathcal{O}\left(\delta_\alpha\right)\times\mathcal{O}\left(\ch^2/k^2\right)\times\left(1+\mathcal{O}\left(\frac{\ch^2}{k^2}\right)\right)\right)\,,
\ee
where $\delta_\alpha$ refers to any of the fluids that are present (concretely, matter and dark energy). 

Since we are interested in scales such that $k\gg\ch$ we can be tempted to neglect the terms of order proportional to $\mathcal{O}\left(\ch^2/k^2\right)$ and $\mathcal{O}\left(\ch^4/k^4\right)$\,, but we need to be careful because we have already seen in Section \ref{matching} that neglecting a term of order $\ch^2/k^2$ in the linear differential system would lead to a wrong result in the limit of pure matter domination. It is only through a numerical solution of the whole system of equations that we can be certain of whether we can leave aside this kind of terms or not. This is actually what we have done, reaching the conclusion that they are actually irrelevant for the redshifts that we are interested in. 

Something similar occurs for the Euler equation. Taking the divergence of \eq{correuler} we see that this equation has the order structure
\begin{align}
\frac{\dot\theta}{\ch^2}&\sim \mathcal{O}(\delta)\times\left(1+\mathcal{O}(\delta)\right)\times\left(\mathcal{O}\left(\frac{\ch^2}{k^2}\right)+\mathcal{O}\left(\frac{k^2}{\ch_s^2}\right)\right)+\mathcal{O}(\delta_\alpha)\times\left(1+\mathcal{O}\left(\frac{\ch^2}{k^2}\right)\right)
\end{align}

Collecting the lowest order terms, we see that $\dot\theta/\ch^2$ is given at the linear level by terms of the order
\be 
\Delta_{E(l)}\sim \mathcal{O}(\delta)\times\left(1+\mathcal{O}\left(\frac{\ch^2}{k^2}\right)+\mathcal{O}\left(\frac{k^2}{\ch_s^2}\right)\right)+\mathcal{O}\left(\delta_\alpha\right)\times\left(1+\mathcal{O}\left(\frac{\ch^2}{k^2}\right)\right)
\ee
At this level, depending on the scale of interest being below or above the sound horizon, the term of order $\delta k^2/\ch_s^2$ may be relevant or not, becoming the dominant one in the limit of very large $k$\,. The same occurs at the non--linear level. 

\section{The time renormalization group} \label{trg}

The time renormalization group (TRG) is a useful technique, introduced in \cite{Pietroni:2008jx}\,, to compute non--linear corrections to the power spectrum and higher order correlators in the framework of Eulerian perturbation theory. The principal idea behind it is the recursive application of the equations for the perturbations.  For any field $\varphi(\mathbf{k},\tau)$ formed by several components we define:
\begin{align}
\ds\langle \varphi_a({\bf k}) \varphi_b({\bf q})\rangle &\equiv \delta_D({\bf k + q}) P_{ab}({\bf k})
\\ \ds\langle \varphi_a({\bf k}) \varphi_b({\bf q})\varphi_c({\bf p})\rangle &\equiv \delta_D({\bf k + q+p})
 B_{abc}({\bf k},\,{\bf q},\,{\bf p})
\\\nonumber \ds\langle \varphi_a({\bf k}) \varphi_b({\bf q})\varphi_c({\bf p})\varphi_d({\bf r})\rangle &\equiv \delta_D({\bf k + p+q+ r}) \,Q_{abcd}({\bf k}\,,{\bf q}\,,{\bf p}\,,{\bf r})\\
\nonumber\,\,&+\delta_D({\bf k + q })\, \delta_D({\bf p + r })P_{ab}({\bf k})P_{cd}({\bf p})
\\\nonumber &\,+\delta_D({\bf k + p}) \,\delta_D({\bf q + r }) P_{ac}({\bf k})P_{bd}({\bf q})
\\\ds &\,+\delta_D({\bf k + r})\, \delta_D({\bf q + p }) P_{ad}({\bf k})P_{bc}({\bf q})\,,
\end{align}
where we have omitted the time dependencies to abbreviate the notation. As usual, $P_{ab}({\bf k})$ is the power spectrum, $B_{abc}({\bf k},\,{\bf q},\,{\bf p})$ the bispectrum, and
$Q_{abcd}({\bf k}\,,{\bf q}\,,{\bf p}\,,{\bf r})$\,, the connected part of the four--point function, the trispectrum. We will neglect the trispectrum in our computations; and therefore the four--point function will be fully given by the power spectrum.

If we define 
\be 
\varphi^t=e^{-\eta}\left(\delta_m\,,\,-\theta_m/\ch\,,\,\delta_x\,,\,-\theta_x/\ch\right)\,,
\ee
where the superindex `$\,^t\,$' simply denotes matrix transposition and
\be \label{etadef}
\eta=\log\frac{a}{a_{in}}\,,
\ee
we can write the continuity and Euler equations for matter and dark energy in the following form
\be
\varphi_a^\prime({\bf k}, \eta)= -\Omega_{ab}({\bf k},\,\eta )
\varphi_b({\bf k}, \eta) + e^\eta \int d^3\mathbf{p}\,d^3\mathbf{q}\,\delta_D(\mathbf{k}-\mathbf{p}-\mathbf{q})
\gamma_{abc}({\bf k},\,-{\bf p},\,-{\bf q},\,\eta )  
\varphi_b({\bf p}, \eta )\,\varphi_c({\bf q}, \eta )\,,
\label{compact}
\ee
where primes denote derivatives with respect to $\eta$\,. The linear evolution of the perturbations is given by the matrix $\Omega_{ab}({\bf k},\,\eta )$\, and the non--linear contributions come from the vertices $\gamma_{abc}\left(\mathbf{k},\mathbf{p},\mathbf{q},\,\eta \right)$\,. 

Recursive iteration of the equation \eq{compact} gives
\begin{align} 
 \partial_\eta\,\langle \varphi_a \varphi_b\rangle &= -\Omega_{ac} 
\langle \varphi_c \varphi_b\rangle   - 
\Omega_{bc} 
\langle \varphi_a \varphi_c\rangle +e^\eta \gamma_{acd}\langle \varphi_c\varphi_d \varphi_b\rangle +e^\eta \gamma_{bcd}\langle \varphi_a\varphi_c \varphi_d\rangle\,,
\\ \partial_\eta\,\langle \varphi_a \varphi_b  \varphi_c \rangle & =  -\Omega_{ad} 
\langle \varphi_d \varphi_b\varphi_c\rangle  -\Omega_{bd} 
\langle \varphi_a \varphi_d\varphi_c\rangle  -\Omega_{cd} 
\langle \varphi_a \varphi_b\varphi_d\rangle
\nonumber\\
&+e^\eta \gamma_{ade}\langle \varphi_d\varphi_e \varphi_b\varphi_c\rangle  
+e^\eta \gamma_{bde}\langle \varphi_a\varphi_d \varphi_e\varphi_c\rangle +e^\eta \gamma_{cde}\langle \varphi_a\varphi_b \varphi_d\varphi_e\rangle \,,
\label{tower}
\end{align}
where we have omitted the momentum dependencies because they can be tracked with the indices. Using in these two equations, the definitions of the correlation functions that we have introduced above, we get the following equations for the power spectrum and the bispectrum:
\begin{align}\nonumber 
&\ds  \partial_\eta\,P_{ab}({\bf k}) =\\
& - \Omega_{ac} ({\bf k})P_{cb}({\bf k})  - \Omega_{bc} ({\bf k})P_{ac}({\bf k}) \nonumber\\
&+e^\eta \int d^3 q\, \left[ \gamma_{acd}({\bf k},\,{\bf -q},\,{\bf q-k})\,B_{bcd}({\bf k},\,{\bf -q},\,{\bf q-k})+ B_{acd}({\bf k},\,{\bf -q},\,{\bf q-k})\,\gamma_{bcd}({\bf k},\,{\bf -q},\,{\bf q-k})\right]\label{syst1}
\end{align}
\begin{align}
\nonumber 
&\ds  \partial_\eta\,B_{abc}({\bf k},\,{\bf -q},\,{\bf q-k}) =\\&  - \Omega_{ad} ({\bf k})B_{dbc}({\bf k},\,{\bf -q},\,{\bf q-k})- \Omega_{bd} ({\bf -q})B_{adc}({\bf k},\,{\bf -q},\,{\bf q-k})\nonumber- \Omega_{cd} ({\bf q-k})B_{abd}({\bf k},\,{\bf -q},\,{\bf q-k})\\&+ 2 e^\eta \left[ \gamma_{ade}({\bf k},\,{\bf -q},\,{\bf q-k}) P_{db}({\bf q})P_{ec}({\bf k-q})+\gamma_{bde}({\bf -q},\,{\bf q-k},\,{\bf k}) P_{dc}({\bf k-q})P_{ea}({\bf k})\right.\nonumber\\&\quad\quad+\left.\gamma_{cde}({\bf q-k},\,{\bf k},\,{\bf -q}) P_{da}({\bf k})P_{eb}({\bf q})\right]\,,
\label{syst2}
\end{align}
where all the functions are also time dependent and we have used the vertex symmetry \eq{perm} to group terms coming from different correlators in \eq{tower}\,.

The matrix that gives the linear evolution of the perturbations in the case of non--interacting dark matter and dark energy is
\be \label{matrix}
\Omega_{ab}=
\begin{pmatrix} 
1 & -1 & 0 & 0 & \\
-\frac{3}{2}\Omega_m & 2+\frac{\ch'}{\ch}+\frac{9}{2}\Omega_m\frac{\ch^2}{k^2} & -\frac{3}{2}\Omega_x & \frac{9}{2}(1+w)\Omega_x\frac{\ch^2}{k^2} \\
0 & 0 & 1-3w+3\cshat & -(1+w)\left(1+9\left(\cshat-w\right)\frac{\ch^2}{k^2}\right) & \\
-\frac{3}{2}\Omega_m & \frac{9}{2}\Omega_m\frac{\ch^2}{k^2}  & -\frac{3}{2}\Omega_x+\frac{\cshat k^2}{(1+w)\ch^2} &2+\frac{\ch'}{\ch}-3\cshat+\frac{9}{2}(1+w)\Omega_x\frac{\ch^2}{k^2}
\end{pmatrix}\,,
\ee
according to \eq{compact} and the equations \eq{eqc} and \eq{eqe}. The non--linear part of the evolution is determined by the vertices 
\be 
\gamma_{abc}\left(\mathbf{k},\mathbf{p},\mathbf{q}\right)=\delta_D\left(\mathbf{k}+\mathbf{p}+\mathbf{q}\right)\Gamma_{abc}\left(\mathbf{k},\mathbf{p},\mathbf{q}\right)\,,
\ee
which, by construction, have the permutation symmetry
\be \label{perm}
\Gamma_{abc}\left(\mathbf{k},\mathbf{p},\mathbf{q}\right)=\Gamma_{acb}\left(\mathbf{k},\mathbf{q},\mathbf{p}\right)
\ee
The ones that correspond to the terms that we have written explicitly in \eq{eqc} and \eq{eqe} are:
\begin{align}
\Gamma_{121}(\mathbf{k},\mathbf{p},\mathbf{q})=\frac{1}{2}\alpha(\mathbf{p},\mathbf{q})\,,\quad \quad 
\Gamma_{222}(\mathbf{k},\mathbf{p},\mathbf{q})=\beta(\mathbf{p},\mathbf{q})\,,
\end{align}
which are the usual ones for matter perturbations.

The equations \eq{syst1} and \eq{syst2} can be simplified for cosmologies for which the matrix $\Omega_{ab}$ is scale independent, becoming:
\begin{align} \label{trt1}
\ds  \partial_\eta\,P_{ab}(k) &= - \Omega_{ac} P_{cb}(k)  - \Omega_{bc} P_{ac}( k)+ e^\eta \frac{4 \pi}{k} \left[I_{acd,bcd}(k)+I_{bcd,acd}(k) \right]
\\ \label{trt2}
\ds  \partial_\eta\,I_{acd,bef}(k) &=- \Omega_{bg} I_{acd,gef}(k) - \Omega_{eg}  I_{acd,bgf}(k)- \Omega_{fg} I_{acd,beg}(k) +2 e^\eta A_{acd,bef}(k)\,,
\end{align}
where we define
\begin{align}
I_{acd,bef}(k) &\equiv \int_{k/2}^\infty dq \,q \int_{|q-k|}^q dp \,p\, \frac{1}{2} \left[ \tilde{\gamma}_{acd}(k, q, p)\, \tilde{B}_{bef}(k,q,p) + (q\leftrightarrow p) \right]\\
A_{acd,bef}(k) &\equiv  \int_{k/2}^\infty dq \,q \int_{|q-k|}^q dp \,p\, \frac{1}{2} \left\{ \tilde{\gamma}_{acd}(k, q, p) \left[ \tilde{\gamma}_{bgh}(k, q, p)P_{ge}(q) P_{hf}(p)\right.\right. 
\nonumber\\
&\left.\left. + \,\tilde{\gamma}_{egh}(q,p,k)P_{gf}(p) P_{hb}(k) +\tilde{\gamma}_{fgh}(p,k,q)P_{gb}(k) P_{he}(q)\right] +
(q\leftrightarrow p)\right\}\, \label{aeq}
\end{align}
with
\begin{align}
 \tilde{\gamma}_{abc}(k, q, p) = \left. \gamma_{abc}({\bf k}, {\bf q}, {\bf p})\right|_{{\bf p}=-({\bf k}+{\bf q})}
\end{align}
and analogously for $\tilde{B}_{abc}(k, q, p)$\,.

 Clearly, $\Omega_{ab}$ is not scale independent for the case of a non--vanishing sound speed of dark energy, but however we can actually use \eq{trt1} and \eq{trt2} reliably, simply because the error introduced by doing so is small. To check this, we fixed the momenta in the $\Omega_{ab}$ matrices appearing in the second and third terms of \eq{trt2} to different values below the Hubble horizon. We found that the solutions are nearly independent on these choices, introducing a relative error between different assignments that is well below $1\%$\,.

The TRG has already been used to study not only the $\Lambda\text{CDM}$ model \cite{Pietroni:2008jx}\,, but also the effects of massive neutrinos \cite{Lesgourgues:2009am} and to take into account high order correlation functions \cite{Anselmi:2010fs}\,. It was also modified for the case of non--Gaussian initial conditions \cite{Bartolo:2009rb}\,, applied to study neutrino quintessence \cite{Brouzakis:2010md}\,, the matter power spectrum for a variable equation of state \cite{Brouzakis:2010hp}\,, the case of coupled quintessence \cite{Saracco:2009df} and dark matter haloes \cite{Elia:2010en}\,. Here we use it to describe the non--linearities in the clustering of matter in the presence of dark energy perturbations.

\section{An approximate numerical solution} \label{appsol}
Solving the full TRG of Section \ref{trg}\,, where the field $\varphi$ has four components and new vertices need to be introduced, requires a large computing time, even if the equations for scale free linear propagation, \eq{syst1} and \eq{syst2}\,, are employed. One can lessen the computational cost considerably by reducing the number of components and, in consequence, of equations. In this section we describe the approximation that we choose in order to achieve it. The equation \eq{supsound} implies that $\delta_x\ll \delta_m$ in matter domination, but this is also true later in the evolution of the universe. As we have already explained, there is roughly one order of magnitude between the matter and dark energy perturbations and therefore $\delta_x$ is always linear on interesting scales, provided that $\delta_m$ remains sufficiently small. Neglecting non--linear terms in the dark energy fluctuations and using the Poisson equation \eq{psn2}\,, we obtain the following continuity and Euler equations for matter perturbations:
\begin{align} \label{2fmcont}
\dot\delta_m(\mathbf{k})&+\theta_m(\mathbf{k})+\int d^3\mathbf{p}\,d^3\mathbf{q}\,\delta_D(\mathbf{k}-\mathbf{p}-\mathbf{q})\alpha(\mathbf{q},\mathbf{p})\theta_m(\mathbf{q})\delta_m(\mathbf{p})=0\\ \label{2fmeul}
\dot\theta_m(\mathbf{k})&+\ch\theta_m(\mathbf{k})+\frac{3}{2}\ch^2\Omega_m\delta_m\left(1+\frac{\Omega_x\delta_x^L}{\Omega_m\delta_m^L}\right)+\int d^3\mathbf{p}\,d^3\mathbf{q}\,\delta_D(\mathbf{k}-\mathbf{p}-\mathbf{q})\beta(\mathbf{q},\mathbf{p})\theta_m(\mathbf{q})\theta_m(\mathbf{p})=0\,,
\end{align}
where $\delta_x^L$ and $\delta_m^L$ are computed from the linear Newtonian theory \eq{linearcontnew} and \eq{lineareulernew} and we have done the further approximation of replacing $\delta_m$ by $\delta_m^L$ in the denominator of the third term in the Euler equation. This allows us to reduce the number of components of the $\varphi$ field to the two matter variables:
\be \label{phim}
\varphi_{m}^t=e^{-\eta}\left(\delta_m\,,\,-\theta_m/\ch\right)
\ee
with 
\be \label{matrixm}
\Omega_{ab}^{(m)}({\bf k},\,\eta )=
\begin{pmatrix} 
1 & -1 & \\
-\frac{3}{2}\Omega_m\left[1+\left(\Omega_x\delta_x^L\right)/\left(\Omega_m\delta_m^L\right)\right] & 2+\frac{\ch'}{\ch} &
\end{pmatrix}\,,
\ee
where we have neglected the subleading terms of the Poisson equation. We have checked numerically that at late times (i.e. for small redshifts) the effect of adding those terms is irrelevant for scales $k\gtrsim 0.01\,h\, \text{Mpc}^{-1}$\,. Notice that the only vertices that we need in this approximation: $\gamma_{112}$ and $\gamma_{222}$\,, correspond to the usual ones for matter (see \cite{Pietroni:2008jx}, for instance)\,. This is the kind of approximation that was already advocated in \cite{Pietroni:2008jx} (and later used in \cite{Lesgourgues:2009am} for studying the effect of massive neutrinos)\,. 

\section{Numerical Results} \label{numerical}

Here we present the results of solving the Newtonian approximation for our system of dark matter and clustering dark energy applying the TRG. The main observable we are interested in is the two point correlation function (the power spectrum). We obtain the matter power spectrum $P_m=\langle\delta_m^2\rangle$ and the total power spectrum $P_{tot}(\mathbf k)$ defined as 
\begin{align} \label{psapp}
P_{tot}(\mathbf{k})=\Omega_m^2 \langle \delta_m(\mathbf{k})^2 \rangle + 2 \Omega_m\Omega_x\langle \delta_m(\mathbf{k})\delta_x(\mathbf{k}) \rangle + \Omega_x^2\langle \delta_x(\mathbf{k})^2 \rangle
\end{align}
Besides, we present results for the (total and matter) growth functions and matter growth index, which we define in Subsection \ref{growths}\,. The total power spectrum $P_{tot}(\mathbf{k})$ is related to the total relative energy density perturbation of dark matter and dark energy, that we define to be:
\begin{align} \label{tcd}
\delta_{tot}=\Omega_m\delta_m+\Omega_x\delta_x
\end{align}
The total clustering density perturbation is $\delta\rho_{tot}=\delta\rho_m+\delta\rho_x$\,. This is the density perturbation that appears in the Poisson equation \eq{poisson} and therefore the one that affects the dynamics of clustering and structure formation. Since we want to confront this density perturbation with the total energy budget, we divide it by the total background density $\bar\rho_{tot}=\bar\rho_m+\bar\rho_x$\,, obtaining the total relative clustering energy density perturbation \eq{tcd}\,, which is the relevant quantity in \eq{psn2}\,. In the case of $\Lambda$CDM\,, dark energy does not cluster because $\Lambda$ is perfectly homogeneous and isotropic. This means that the total relative clustering energy density perturbation for $\Lambda$CDM is $\Omega_m\delta_m$\,.

We choose the cosmological parameters to be close to those of the current best fits, assuming a flat universe. In particular, we take $\Omega_{m}^0=0.25$\,, $\Omega_{b}^0h^2=0.0224$\,, $h=0.72$\,, 
$n=0.97$ and $\sigma_{8}=0.8$ (defined under the assumption of $w=-1$\,)\,. Other reasonable choices do not change our conclusions in any respect. The initial time, $\eta=0$\,, is associated to 
redshift  $z_{in}=100$\,. At this time, we set the initial conditions for the TRG equations according to the relation between the dark energy perturbations and the cold dark matter ones given by the equations \eq{supsound} and \eq{velrel}\,. As input, we introduce the linear cold dark matter power spectrum obtained from CAMB \cite{Lewis:1999bs}\,. To run the TRG integration we choose the initial values for the bispectra equal to zero as we neglect all the non--Gaussianities generated at redshifts higher than $z=100$\,. Recall that within our approximation the trispectrum remains zero at all times. 

Let us now describe the two different power spectra that we calculate. In Subsection \ref{growths} we plot the matter and total growths and explain the non--linear effect on the matter growth index.

\begin{figure}
\begin{center}
 \subfloat{\includegraphics[width=0.6\textwidth]{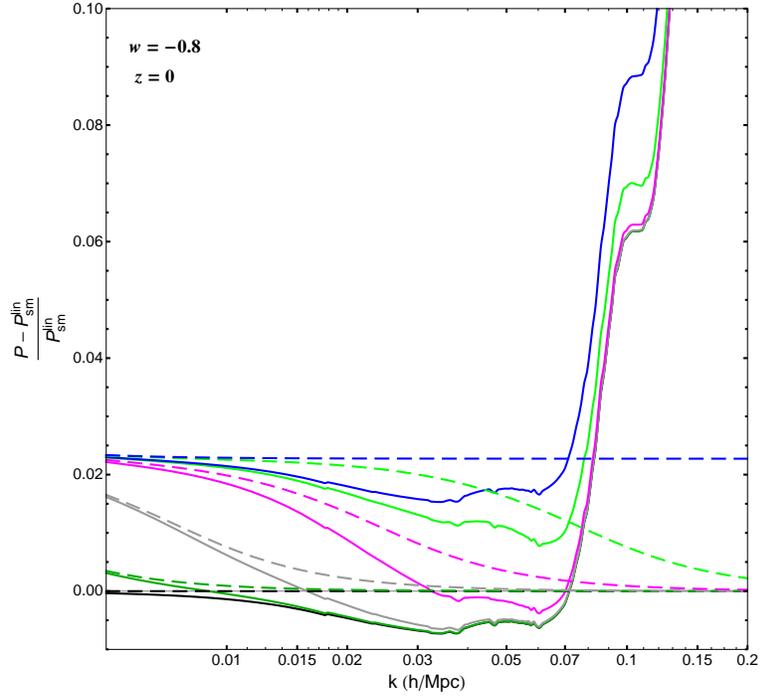}}
 \vspace{1.5cm}
 \subfloat {\includegraphics[width=0.6\textwidth]{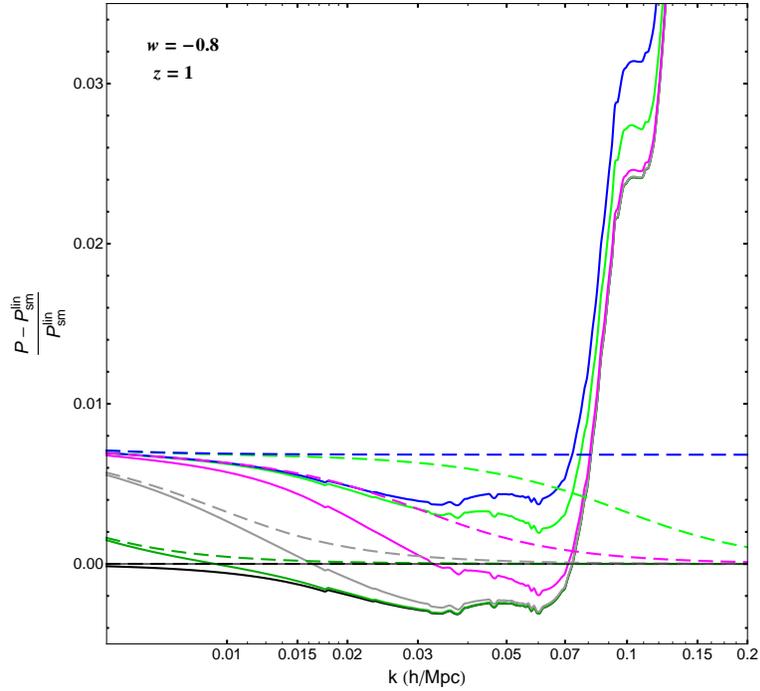}}
\caption{\small{Relative difference between the matter spectra and their linear smooth counterpart. The spectra are evaluated for $w=-0.8$ at redshifts $z=0$ (top) and $z=1$ (bottom). Dashed lines correspond to linear spectra while solid ones are non--linear ones computed using the TRG. The black lines represent the smooth case while the dark green ($\cshat=0.1$), grey ($0.01$), magenta ($0.001$), light green ($0.0001$) and blue (0) show the results for various sound speeds of dark energy.}}
\label{PSmatterw08}
\end{center}
\end{figure}

\begin{figure}
\begin{center}
 \subfloat{\includegraphics[width=0.6\textwidth]{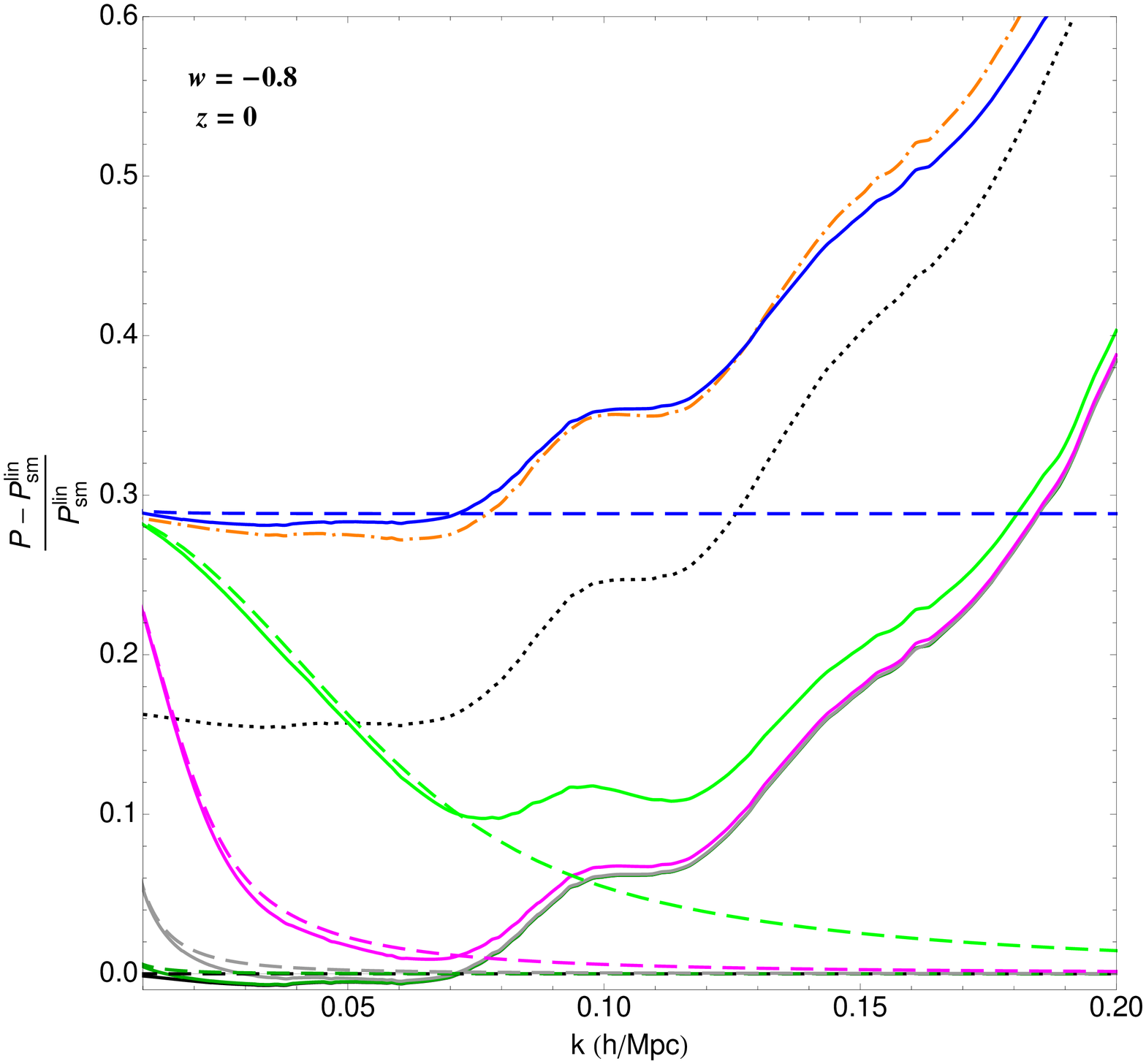}}
 \vspace{1.5cm}
 \subfloat {\includegraphics[width=0.6\textwidth]{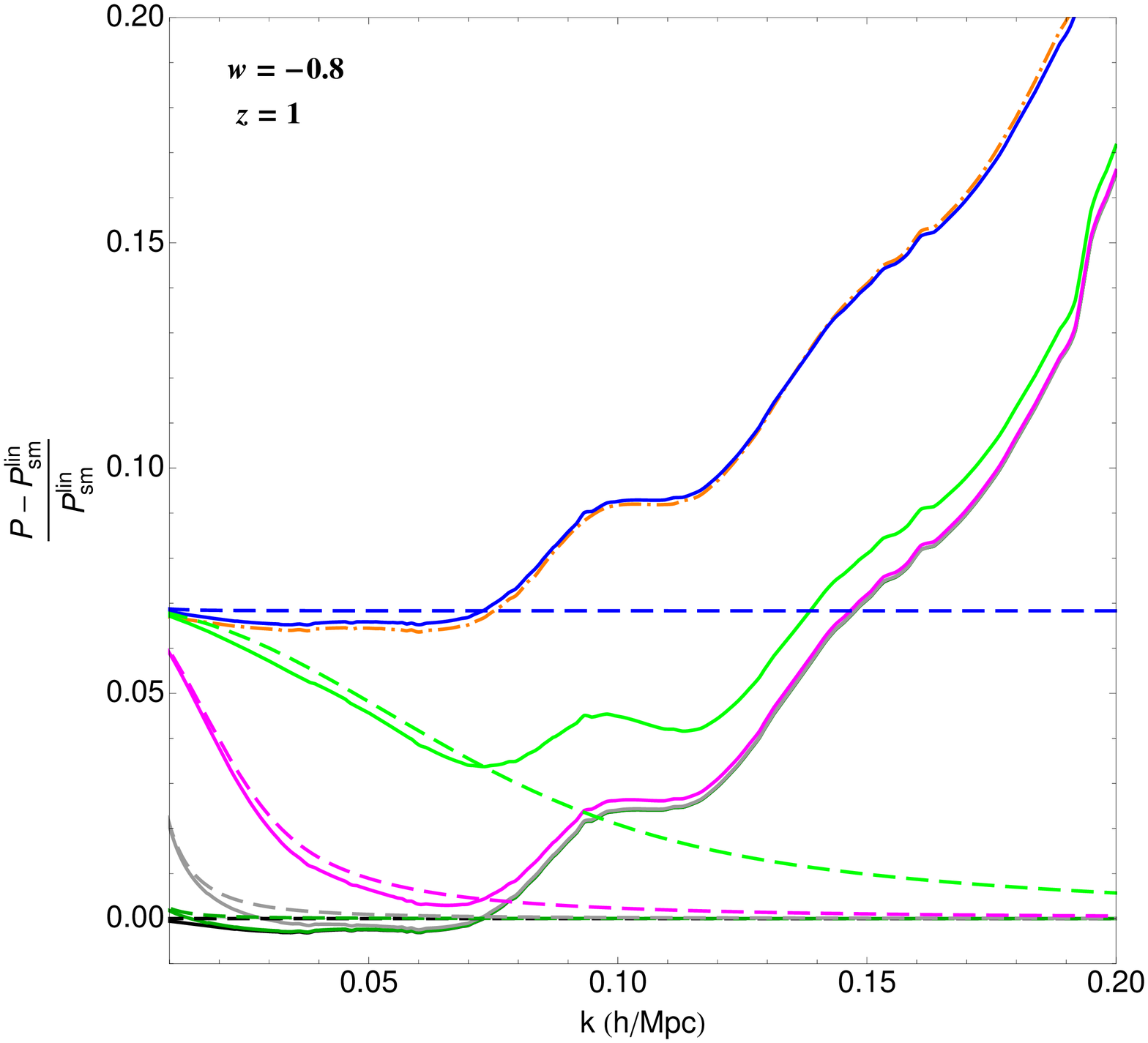}}
\caption{\small{Relative difference between the total power spectrum (matter and dark energy, defined as in \eq{psapp})  and its linear smooth counterpart. The spectra are evaluated for $w=-0.8$ at redshifts $z=0$ (top) and $z=1$ (bottom). Dashed lines correspond to linear spectra while solid ones are non--linear results computed using the TRG. The black lines represent the smooth case while the dark green ($\cshat=0.1$), grey ($0.01$), magenta ($0.001$), light green ($0.0001$) and blue (0) show the results for various sound speeds of dark energy. The black dotted line corresponds to the $\Lambda \text{CDM}$ cosmology and the orange dash--dotted one to the numerical approximation of the Appendix for $\cshat=0$\,.}}
\label{PStotw08}
\end{center}
\end{figure}

\subsection{The power spectra}

We compute the linear $P_m$ solving the continuity \eq{linearcontnew} and Euler \eq{lineareulernew} equations for matter and dark energy, which can be expressed in terms of the matrix \eq{matrix}\,. In practice, we work with the Poisson equation \eq{psn2} because, as we have already explained, using the full Poisson expression for $\phi_N$ including the velocity perturbations does not alter the results. We also calculate the matter power spectrum non--linearly using the TRG, applying the equations \eq{trt1} and \eq{trt2} with the approximation described in the previous section (in which we treat linearly the contribution from dark energy perturbations)\,. Comparing the two results we observe and describe the non--linear effects on the matter power spectrum.

We proceed in the same way for the total clustering spectrum $P_{tot}$ of matter and dark energy. The linear version of it is given by \eq{psapp} where $\delta_m$ and $\delta_x$ are both computed at the linear level. Our approximation to its non--linear counterpart is described in Section \ref{appsol}\,.

In the particular case of zero sound speed of dark energy, the equations for the velocity perturbations of matter and dark energy are identical. For the sake of comparison, we also compute the total non--linear power spectrum making use of this fact to simplify the system of equations, as explained in the Appendix.

Let us notice that, strictly speaking, in order to compute a power spectrum at second order we should know the density perturbations at order three. This is because if we take into account contributions of order $\delta^{(2)}\delta^{(2)}$ we should also include contributions of the type $\delta^{(1)}\delta^{(3)}$\,, where the number in parenthesis includes the perturbation order. Clearly, these two types of contributions are the smallest of the ones we need to consider and therefore we can work (as we do) expanding the continuity and Euler equations up to order two in the perturbations.

\subsubsection{Matter power spectrum} \label{mpss}

\begin{figure}
\begin{center}
 \subfloat{\includegraphics[width=0.6\textwidth]{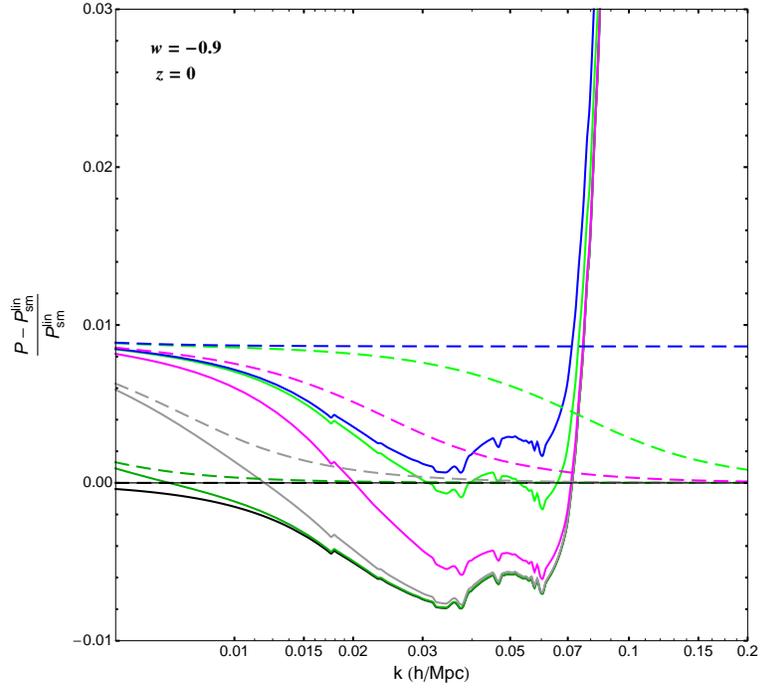}}
 \vspace{1.5cm}
 \subfloat {\includegraphics[width=0.6\textwidth]{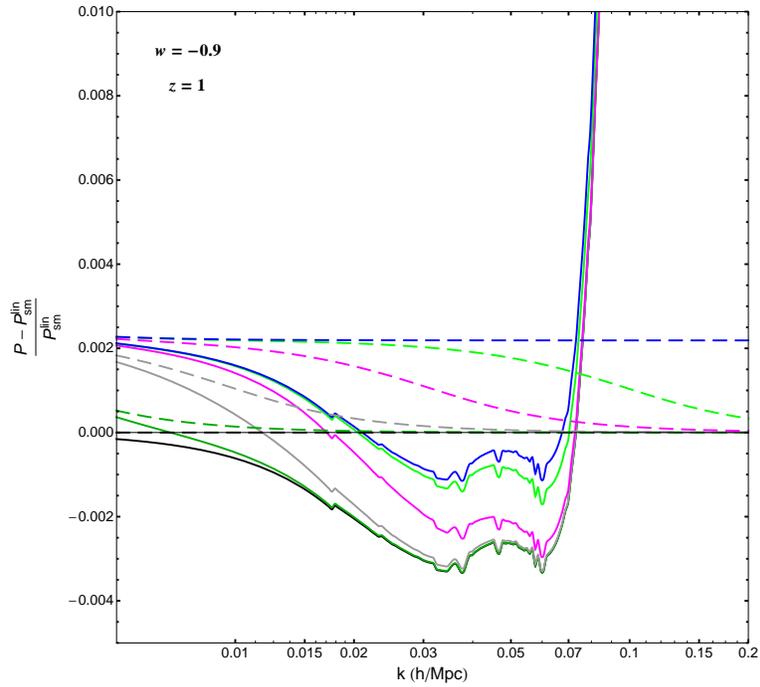}}
\caption{\small{Relative difference between the matter spectra and their linear smooth counterpart. The spectra are evaluated for $w=-0.9$ at redshifts $z=0$ (top) and $z=1$ (bottom). Dashed lines correspond to linear spectra while solid ones are non--linear ones computed using the TRG. The black lines represent the smooth case while the dark green ($\cshat=0.1$), grey ($0.01$), magenta ($0.001$), light green ($0.0001$) and blue (0) show the results for various sound speeds of dark energy.}}
\label{PSmatterw09}
\end{center}
\end{figure}

The Figures \ref{PSmatterw08} and \ref{PSmatterw09} correspond to the matter power spectra for two values of the equation of state of dark energy: $w=-0.8$ and $-0.9$ respectively. Each curve corresponds to a power spectrum either linear (dashed lines) or non--linear (continuous lines). The figures show the relative difference between a spectrum and its linear smooth counterpart $P_{\text{sm}}^{\text{lin}}$\,, which is defined as the matter linear power spectrum in the limit of $\cshat= 1$ and, in practice, can be calculated from \eq{matrixm} setting $\delta_x^L=0$\,. Clearly, the only parameter that affects $P_{\text{sm}}^{\text{lin}}$ is the equation of state of dark energy (through the background expansion). This spectrum is represented in the figures by the dashed black line, it is scale independent and, of course, sits at zero. The black continuous line shows how $P_{\text{sm}}^{\text{lin}}$ gets modified by the non--linear (pure cold dark matter) corrections to the Euler and continuity equations. We can see that for large scales the difference is minimal but at small ones it grows very fast. For $k=0.1\,h\,\text{Mpc}^{-1}$ it is of order 6$\%$ for redshift zero in the case of $w=-0.8$\,. 

The different colors of the curves distinguish between values of the sound speed of dark energy. The differences between the dashed lines (linear spectra) in each of the Figures \ref{PSmatterw08} and \ref{PSmatterw09} are due to the enhancement of the matter power spectrum that the dark energy clustering induces at large scales. Clearly, the closest is $\cshat$ to the speed of light the smaller is the effect, which is maximal for $\cshat=0$\,. For scales larger than the sound horizon of dark energy ($k<\ch_s$) the dark energy fluctuations cluster. This contributes to enhance the gravitational potential, which in turn produces an increment of the matter perturbations. The gentle decays that are observed at different scales for different speeds of sound are due to the transition between the regimes $k<\ch_s$ and $k>\ch_s$\,. This effect is similar to that of neutrino free--streaming. It can also be observed in the non--linear spectra (continuous lines) for large scales, but then, once the non--linearities become important (they very clearly do so at around $0.07h\, \text{Mpc}^{-1}$)\,, the linear suppression of the spectra is washed out and more difficult to see. 

The feature that is observed at $0.1h\, \text{Mpc}^{-1}$ in Figure \ref{PSmatterw08} is due to the effect of non-linearities on the valley of the baryon acoustic oscillations that appears to the right of the first peak (which is roughly located at $0.07h\, \text{Mpc}^{-1}$)\,. Similarly, the small bump that appears between $0.04 h\, \text{Mpc}^{-1}$ and $0.06h\, \text{Mpc}^{-1}$ in Figures \ref{PSmatterw08} and \ref{PSmatterw09} corresponds to the same effect on the valley to the left of the first peak. One can understand these features realizing that we plot $1/P_{\text{sm}}^{\text{lin}}$ (and so the valleys become peaks) modified by a factor that measures the coupling between modes of different momenta in the non--linear power spectrum.

A remarkable property of both figures is that the non--linearities seem roughly independent on the sound speed. In reality, the correction is $\cshat$ dependent but it varies at most by approximately $0.5\%$ changing the sound speed between 0 and 1 (for the smallest scales $\sim 0.1\,h\,\text{Mpc}^{-1}$ in the case of $z=0$ and $w=-0.8$\,)\,. The TRG resums corrections at all orders in perturbation theory. The leading ones correspond to 1--loop diagrams that involve the product of two linear power spectra (as can be seen from the pair of equations \eq{syst1} and \eq{syst2}\,) in such a way that the order of the resulting corrections can be estimated in terms of the correction to the power spectrum itself. Concretely, the difference between the linear matter power spectra $P_m$ and $P_{\text{sm}}^{\text{lin}}$ is approximately $P \sim (1+\Delta) P_{\text{sm}}^{\text{lin}}$\,, where $\Delta\sim 2\%$ (at most) for $\cshat=0$ (see Figure \ref{PSmatterw08})\,. Therefore, the leading non--linear corrections from the TRG are of the order $\int  P^2 dk \sim (1+2\Delta)\int dk \left(P_{\text{sm}}^{\text{lin}}\right)^2$\,, where $\int dk$ formally denotes the momentum integrals. We see from Figure \ref{PSmatterw08} that $P/P_{\text{sm}}^{\text{lin}}\sim 1.1$ for $k\sim 0.15\, h\, \text{Mpc}^{-1}$ (which roughly corresponds to the limit of validity of the TRG method). From here we can estimate the maximum non--linear correction due to the sound speed (happening for $\cshat=0$) which turns out to be approximately the $0.5\%$ difference that we mentioned above\,. Clearly, for sound speeds larger than zero the effect is even smaller and so we see that indeed the effect of the sound speed on the non--linearities is very small.

The importance of the non--linearities increases as we reduce the redshift. For instance, if we look at the case $w=-0.8$ and $k=0.1\,h\, \text{Mpc}^{-1}$\,, we see that the $2.4\%$ difference with respect to $P_{\text{sm}}^{\text{lin}}$ observed in the bottom panel ($z=1$) of Figure \ref{PSmatterw08} becomes, as we already said, $6\%$ for redshift 0 (top panel)\,.

\begin{figure}
\begin{center}
 \subfloat{\includegraphics[width=0.6\textwidth]{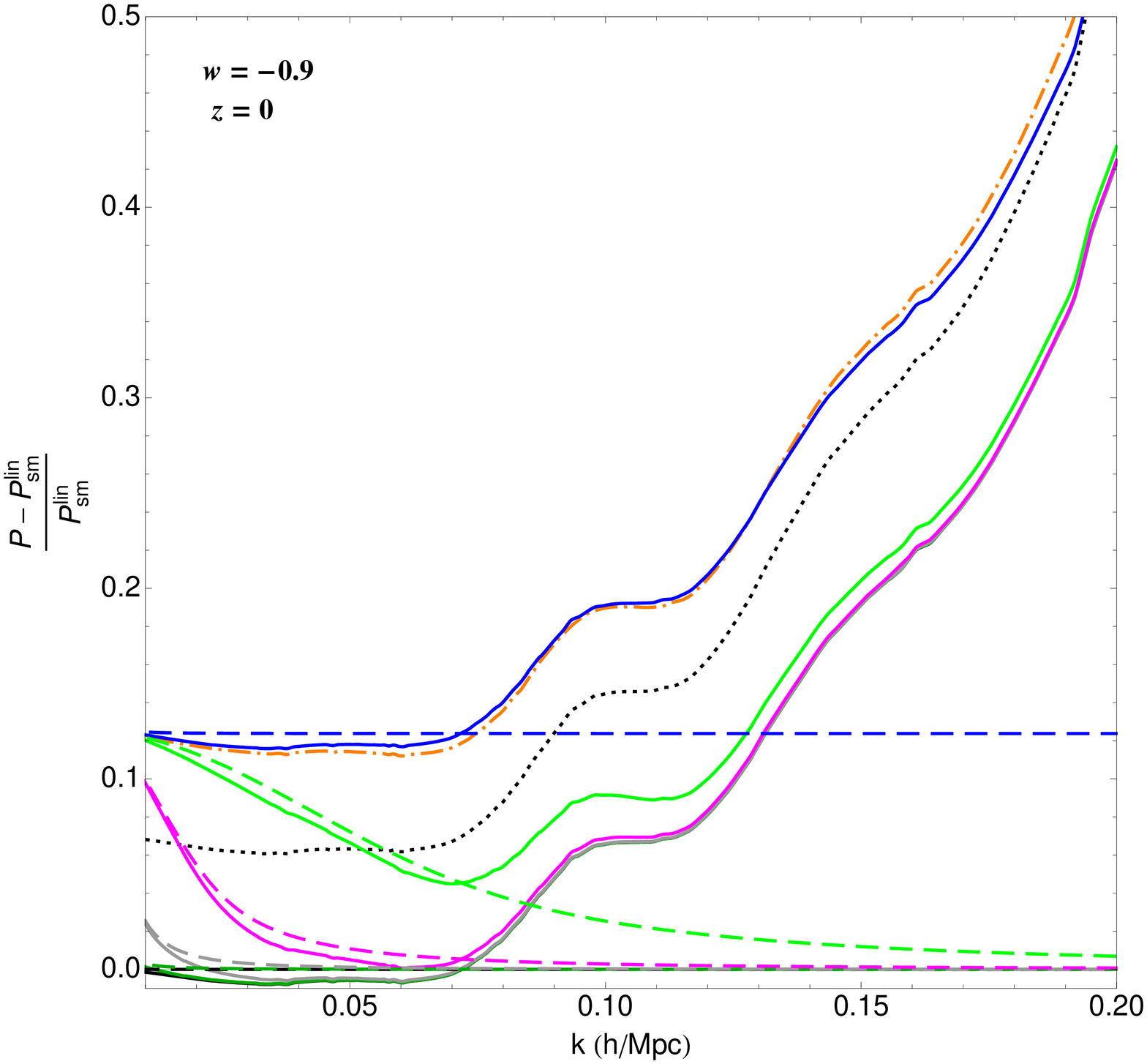}}
 \vspace{1.5cm}
 \subfloat {\includegraphics[width=0.6\textwidth]{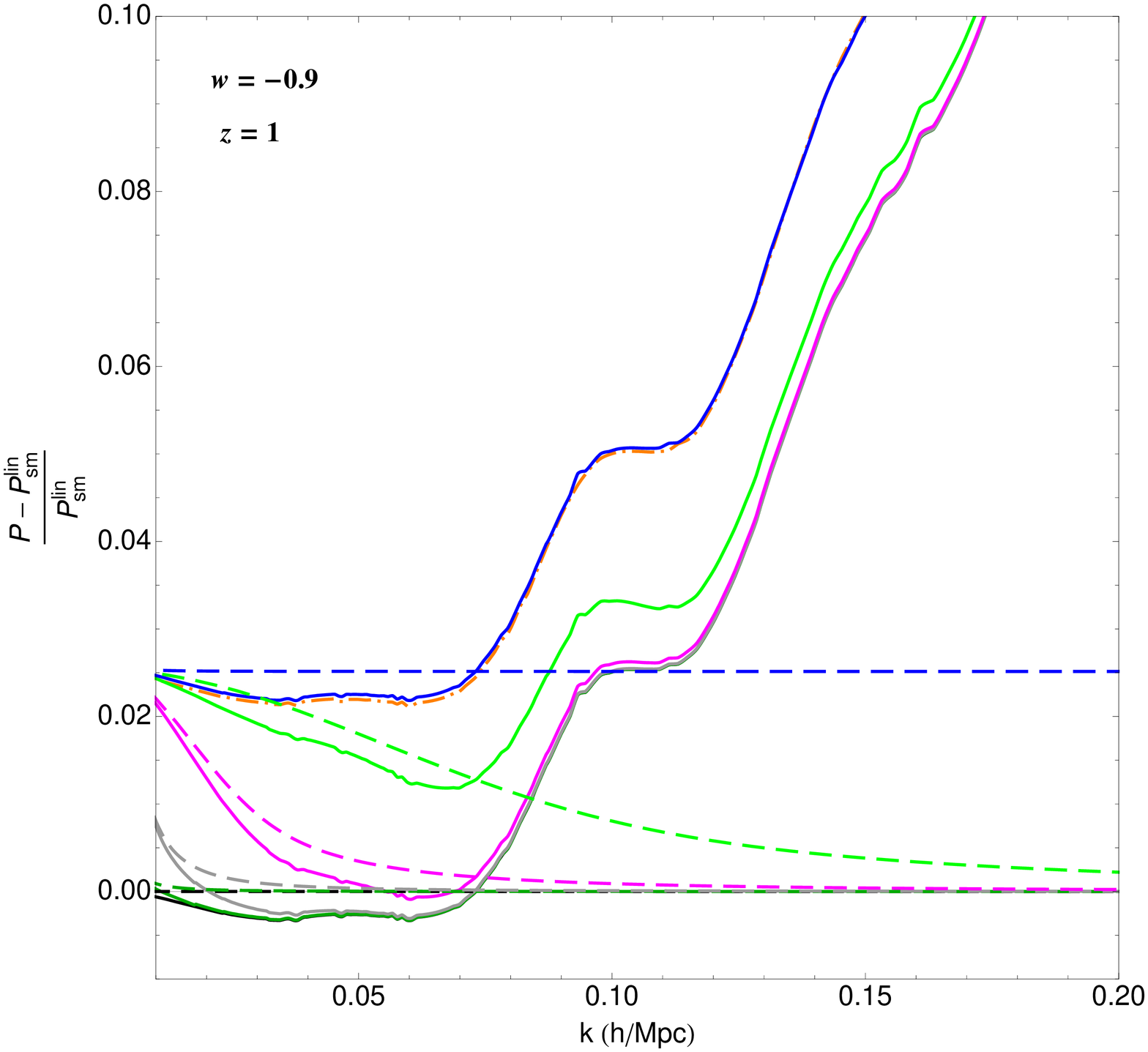}}
\caption{\small{Relative difference between the total power spectrum (matter and dark energy, defined as in \eq{psapp}) and its linear smooth counterpart. The spectra are evaluated for $w=-0.9$ at redshifts $z=0$ (top) and $z=1$ (bottom). Dashed lines correspond to linear spectra while solid ones are non--linear ones computed using the TRG. The black lines represent the smooth case while the dark green ($\cshat=0.1$), grey ($0.01$), magenta ($0.001$), light green ($0.0001$) and blue (0) show the results for various sound speeds of dark energy. The black dotted line corresponds to the $\Lambda \text{CDM}$ cosmology and the orange dash--dotted one to the numerical approximation of the Appendix for $\cshat=0$\,.}}
\label{PStotw09}
\end{center}
\end{figure}

\subsubsection{Total power spectrum}

The Figures \ref{PStotw08} and \ref{PStotw09} correspond to the total power spectrum for $w=-0.8$ and $-0.9$ respectively. As for the matter ones, each curve corresponds to a power spectrum either linear (dashed lines) or non--linear (continuous lines) and we use the same color code as before. Now, we denote by $P_{\text{sm}}^{\text{lin}}$ the linear total power spectrum for $\cshat= 1$ and, in practice, it can be calculated from the linear evolution equations setting to zero the dark energy fluctuations. We see, according to the definition \eq{psapp}\,, that this spectrum has now an $\Omega_m^2$ prefactor that does not appear in the pure matter case. As we already mentioned before, the total power spectrum is computed from \eq{psapp} as explained in Section \ref{appsol}\,. 

There are two extra lines with respect to the matter plots. The orange dash--dotted corresponds to the approximation of the Appendix and the black dotted one to the $\Lambda\text{CDM}$ cosmology. This last one is out of the figures for redshift 1 due to the enhancement from $\Omega_m$\,. 

Following the discussion in Section \ref{appsol}\,, the orange line serves us to confirm the validity of our approximation of treating linearly the dark energy density perturbation $\delta_x$\,. The fact that the difference between the continuous blue line (non--linear, $\cshat=0$\,) and the orange one (the approximation of the Appendix), that corresponds to the single fluid approximation, is so small for any scale means that for sound speeds different from zero, treating $\delta_x$ linearly will be even more justified. The equation \eq{supsound} shows that $\delta_x<\delta_m$ (during pure matter domination, but this will also happen at later epochs). We see that $\delta_x$ can be even an order of magnitude smaller than $\delta_m$ because of the (1+w) factor, therefore the linear approximation can be used for $\delta_x$ even when $\delta_m$ must be treated non--linearly. Besides, the clustering (and hence $\delta_x$) is smaller for larger values of $\cshat$ and therefore, since the plot shows that using the linear equations for the dark energy perturbations works well, we conclude that the linear approximation will be even better for $\cshat$ closer to the speed of light. Although it can not be read directly from the figures, the percentage difference between the spectra calculated as in the Appendix and with our approximation in the case of $\cshat=0$ only becomes bigger than $1\%$ for the largest equation of state that we consider ($w=-0.8$)\,, z=0 and $k\gtrsim 0.16\,h\,\text{Mpc}^{-1}$\,, at the edge of the range of validity of the TRG.

Notice that the feature that appears in the Figures \ref{PSmatterw08} and \ref{PSmatterw09} at $k\sim 0.1\,h\,\text{Mpc}^{-1}$\ is present in the Figures \ref{PStotw08} and \ref{PStotw09} but stretched. The bump at around $k\gtrsim 0.05\,h\,\text{Mpc}^{-1}$ in the matter power spectra can also still be (weakly) seen in Figure \ref{PStotw09} at $k\gtrsim 0.16\,h\,\text{Mpc}^{-1}$, but washed out by the effect of $\Omega_x$ and the choice of a linear horizontal axis.

Looking at the figures \ref{PStotw08} and \ref{PStotw09} for redshift zero, we can see that the difference between the relative $\Lambda\text{CDM}$ and zero sound speed cases depends on the scale. Besides, the redshift dependence is a most interesting feature because it could be a way of detecting a deviation from $\Lambda\text{CDM}$\,, using measurements at a fixed (or several) scales for different redshifts (between 0 and 1, or even a larger range)\,. The idea also applies to the other sound speeds and so this could be a characteristic that may in principle help to discriminate between different values of it.

\subsection{Growth of perturbations}\label{growths}

The growth of matter perturbations
\be \label{growth}
g_m\equiv\frac{\dc}{a}\,, 
\ee
introduced in \cite{Wang:1998gt}\,, is an interesting function to test, for instance, deviations from a $\Lambda\text{CDM}$ cosmology due to an equation of state different from $-1$\,. In the limit of pure matter domination $g_m\rightarrow 1$\,. It is well known that it is possible to fit $g_m$ using a
simple parameterization \cite{Linder:2005in} that defines the matter growth index $\gamma_m$ and depends on the relative energy density of
matter $\Omega_m$\,:
 \be \label{gammadef}
g_m(a)=g_m(a_i)\exp\int_{a_i}^a
\left(\Omega_m(\tilde{a})^{\gamma_{m}}-1\right)\frac{\rmd\tilde{a}}{\tilde{a}}\;.
\ee
A good approximation for the matter growth index in the case of $w\text{CDM}$ cosmologies is \cite{Linder:2005in}
\be
\label{eqlind}
\gamma_m=0.55+0.05\left[1+w(z=1)\right]
\ee
When dark energy perturbations are taken into account (and let us recall that one needs to do so if $w\neq -1$)\,, the growth and the growth index become scale dependent  and vary with the sound speed of dark energy and its equation of state \cite{Ballesteros:2008qk}\,. We can write the equation \eq{gammadef} in a useful way to study these dependencies on the growth index:
\be
\label{eqgamma}
\gamma_m=\left(\log \Omega_m\right)^{-1} \log\left(\frac{\rmd \log \dc}{\rmd \eta}\right)\,,
\ee
where $\eta$ is defined in \eq{etadef}\,. 

\begin{figure}
\begin{center}
\subfloat{\includegraphics[width=1\textwidth]{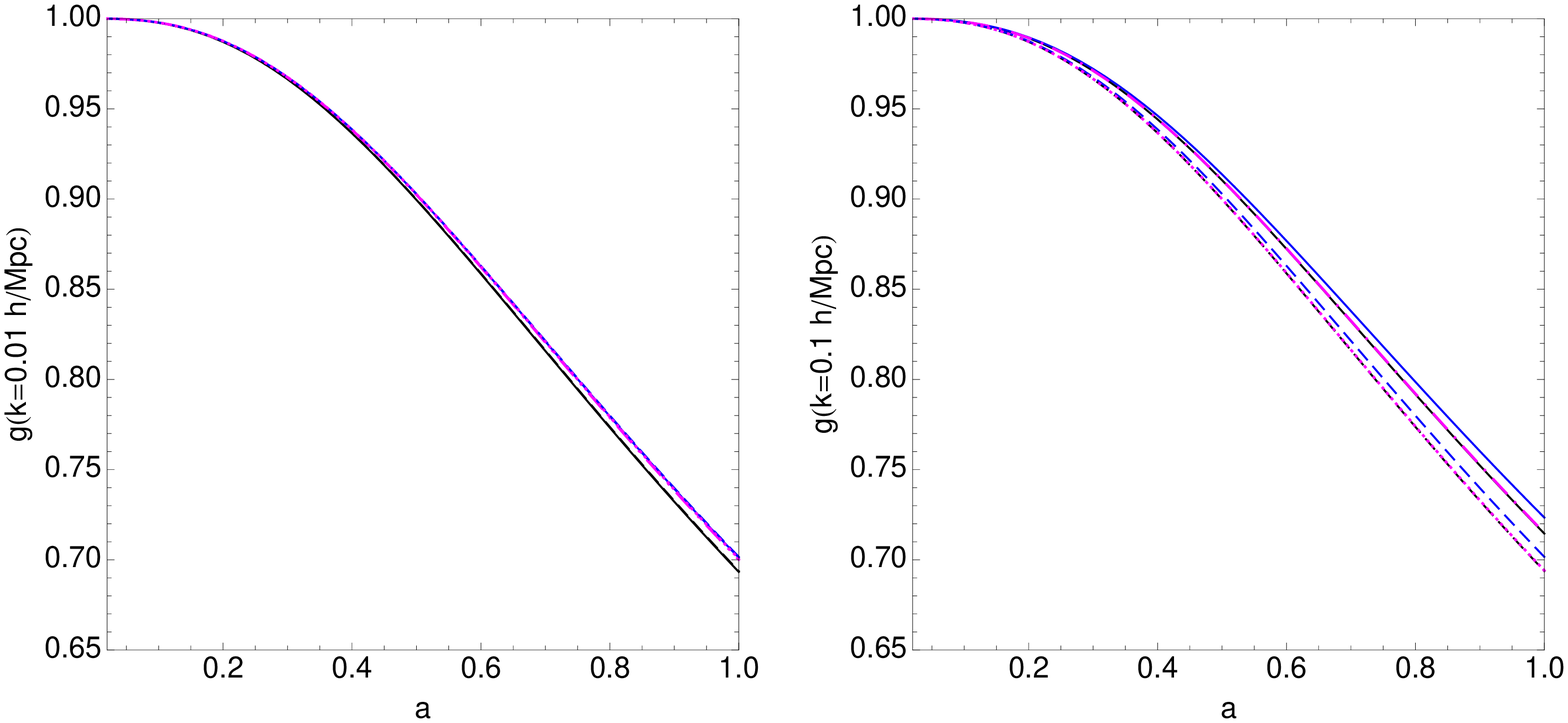}}
\caption{\small{Cold dark matter growth $g_m$ as a function of the scale factor. It is computed for $w=-0.8$ at $k=0.01\,h\,\text{Mpc}^{-1}$ (left) and $k=0.1\,h\,\text{Mpc}^{-1}$ (right)\,. The black dashed curve represents the smooth linear case and the black continuous one the smooth non--linear one.  The magenta ($\cshat=0.001$) and the blue ($0$) show the effect of varying the sound speed of dark energy. The continuous blue is non--linear and the dashed blue is linear. The dash--dotted magenta is non--linear and the dotted is linear. All the non--linear calculations have been done with the TRG.}}
\label{g_w08_matter}
\vspace{1.cm}
\subfloat{\includegraphics[width=1\textwidth]{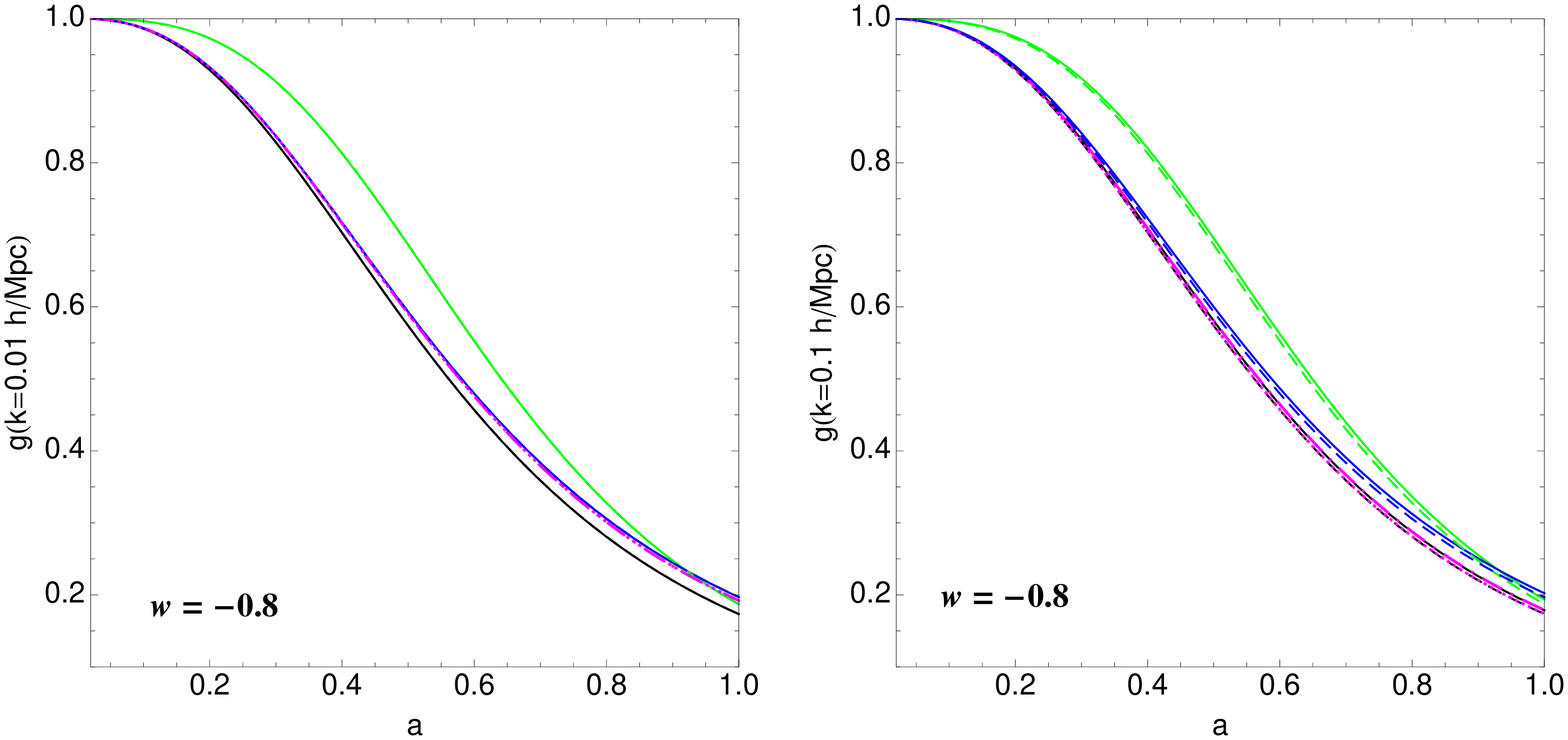}}
\caption{\small{Same as Figure \ref{g_w08_matter} for the total growth $g$ of cold dark matter and dark energy. The green lines represent here the $\Lambda$CDM case, that we have included for comparison. The continuous one is the non--linear calculation and the dashed one corresponds to the linear result.}}
\label{g_w08_tot}
\end{center}
\end{figure}

\begin{figure}[t]
\begin{center}
\includegraphics[width=0.9\textwidth]{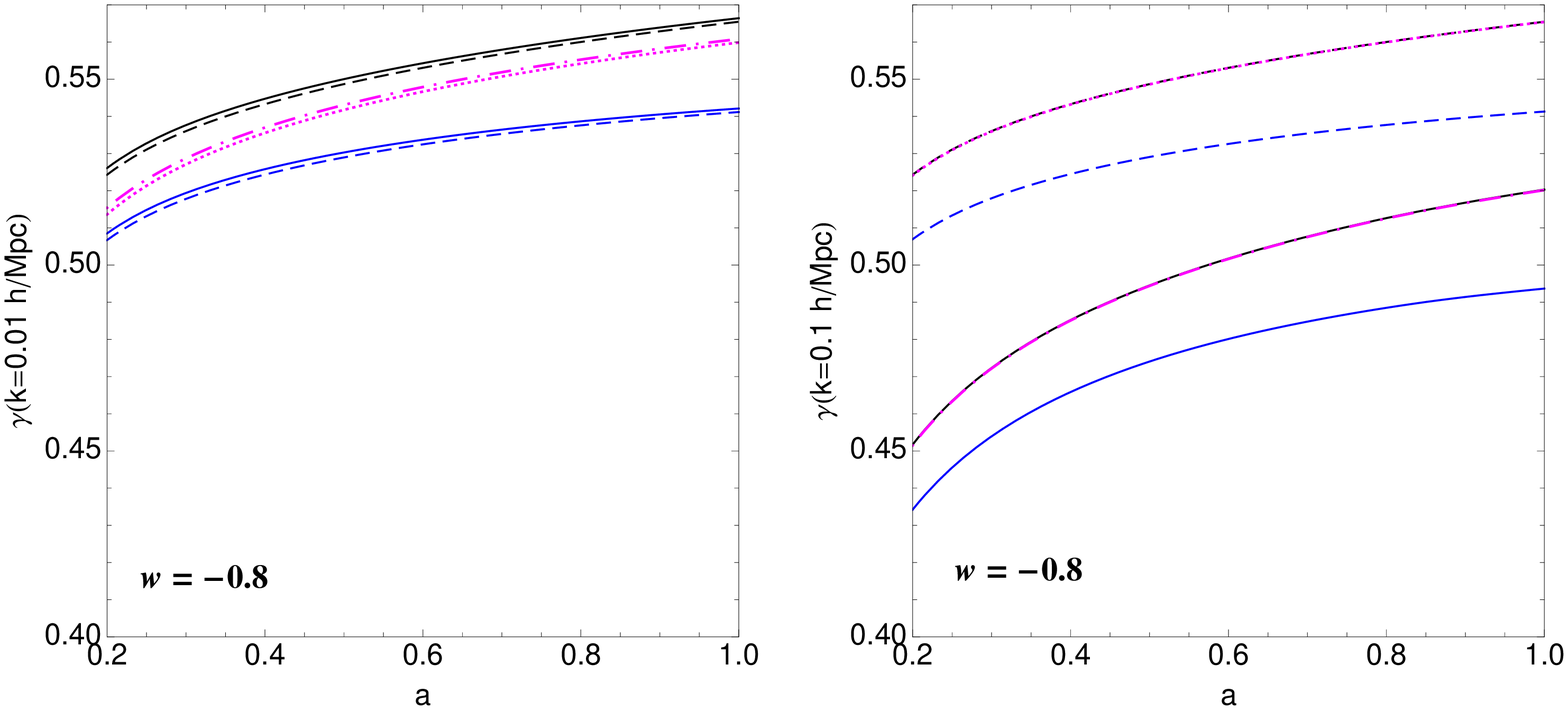}
\caption{\small{Cold dark matter growth index $\gamma_m$ as a function of the scale factor. It is computed for $w=-0.8$ at $k=0.01\,h\,\text{Mpc}^{-1}$ (left) and $k=0.1\,h\,\text{Mpc}^{-1}$ (right)\,. The black dashed curve represents the smooth linear case and the black continuous one the smooth non--linear one.  The magenta ($\cshat=0.001$) and the blue ($0$) show the effect of varying the sound speed of dark energy. The continuous blue is non--linear and the dashed blue is linear. The dash--dotted magenta is non--linear and the dotted is linear. All the non--linear calculations have been done with the TRG.}}
\label{gamma_w08_matter}
\end{center}
\end{figure}

Usually, the growth and growth index are employed at the linear level but it is clear that both quantities receive contributions from the non--linear terms in the evolution equations. In this work we compute these effects taking into account the presence of dark energy perturbations. 

In an analogous way to the one that is customarily done for matter, we can define the total growth function $g_{tot}$ that corresponds to the full clustering perturbation of dark matter and dark energy \eq{tcd}. In order to do this, we simply replace $\delta_m$ in \eq{growth} by $\delta_{tot}=\Omega_m\delta_m+\Omega_x\delta_x$\,. This total growth is relevant for structure formation in situations where the dark energy component clusters. The linear total growth index has been studied in \cite{Sefusatti:2011cm} for the case of zero sound speed.
 
Let us now focus on the cold dark matter growth function which is shown in Figure \ref{g_w08_matter}\,. The smooth linear case is represented by the black dashed curve, which is indistinguishable in the graph from the dotted magenta ($\cshat=0.001$ linear). Let us recall that in the limit $\cshat\rightarrow 1$ we recover the smooth dark energy case. 
The right panel of the Figure \ref{g_w08_matter} shows that the effect of varying the sound speed of dark energy is mainly a linear effect, because it is essentially the same for the linear and non--linear curves. This effect is indeed due to the $\cshat k^2$ term of the (linear) Euler equation. There we see that the blue line (either linear or not) is always above the corresponding magenta one because a smaller sound speed enhances the growth of the dark energy fluctuations and this is communicated to the matter ones through the gravitational potential. Moreover, the right panel shows that for a given sound speed, the non--linear curve is above the linear one because the non--linearities enhance the clustering for small scales. This can also be seen in the figures of the previous section for the matter perturbations. On the other hand, for $k=0.01 h\, \text{Mpc}^{-1}$ (left panel) the blue and magenta curves cannot be distinguished because the effect of the sound speed is very small at large scales.

We already commented in the Introduction that the change in the growth function of matter induced by dark energy perturbations at the linear level can be at most of order $1\%$ under variations of the sound speed. We see from Figure \ref{g_w08_matter} that this is still the case at the non--linear level. However, the non--linear corrections on their own have an even somewhat larger effect ($\sim 3.5\%$) for small scales ($ 0.1\,h\, \text{Mpc}^{-1}$) and a value of $w=-0.8$\,.

Notice that we extract the non--linear $\delta_m$ from the non--linear matter power spectrum computed with the TRG; just by taking the square root of it. This is an approximation with a very small error that peaks for small scales and that we can neglect.

The Figure \ref{g_w08_tot} displays the total growth function for cold dark matter and dark energy. The effects that we have described for matter also hold in this case. The linear total growth function can be easily computed by solving the linear Euler and continuity equations to get the total perturbation $\delta_{tot}=\Omega_m\delta_m+\Omega_x\delta_x$\,. For the non--linear case, let us remind that in our approximation, we actually treat $\delta_x$ linearly.

The Figure \ref{g_w08_tot} tells us that the contribution of dark energy perturbations to the total growth is relevant. It induces a difference with respect to the $\Lambda$CDM case (where dark energy perturbations are absent by definition) that can be as large as $\sim 15\%$ for $a\sim 0.4-0.5$\,. This result is roughly independent on whether the comparison includes the non--linearities or not and the scale at which we look. Interestingly, at zero redshift, the effect of changing the sound speed from zero to one is also of roughly $\sim 15\%$ for $k=0.1\,h\,\text{Mpc}^{-1}$\,.

For completeness, we also provide the matter growth index computed from \eq{eqgamma} in Figure \ref{gamma_w08_matter}\,. We observe that the effect of the non--linearities is negligible for large (linear) scales and becomes of the order of approximately $10-15\%$ at small (non--linear) scales. As we can see, this variation depends mildly on the redshift (for low values of it: $z\sim 0-1$)\,. This is an important result because it means that non--linearities should be taken into account when using the growth index for model discrimination or parameter estimation. Finally, let us remark that this result is essentially independent on the sound speed of dark energy, as it can be checked comparing the relative heights of the curves in Figure \ref{gamma_w08_matter}\,.

\section{Conclusions}\label{conclusions}

The near future will likely bring to us a wealth of information, and 
constraints, on the physical origin of the acceleration of the universe, 
which may be due to dark energy or some modification of general 
relativity. Since at the moment we have almost no clue, neither from 
observations nor from theory, about the possible nature of dark energy, 
it is a sensible attitude to parametrize it in a general way. If dark energy is described as a fluid with a certain equation of state different from $-1$\,,
we can study it at the perturbation level using an arbitrary sound speed and anisotropic stress.
In this work, we focus on the effect of the speed of sound (which must be non--adiabatic)\,. An immediate consequence of a sound speed of dark energy different from the one of light
is the emergence of a new length scale, the dark energy sound horizon, 
leading to an effectively scale dependent growth function for dark 
matter, a feature commonly associated to modified gravity scenarios. 

We have explored the consequences of a non--relativistic speed of sound 
for dark energy, both at the linear and the non--linear levels, emphasizing the need of using the rest frame sound speed.
An accurate treatment of non--linearities is nowadays widely acknowledged
to be crucial in order to compare theory with future galaxy surveys or 
measurements of cosmic shear. The traditional third order Eulerian 
perturbation theory is known to fail at redshifts smaller than unity, so 
that the use of resummation procedures, such as the TRG, that can help to extend 
the validity of  semi--analytical methods to smaller scales and lower 
redshifts is mandatory. 

N--body simulations represent a solidly 
established alternative. However, if one is ultimately interested in (for example) parameter forecasting, numerically faster tools, able to scan 
more efficiently over a multidimensional parameter space, or over 
different models, are definitely useful. For this kind of analysis, semi--analytical methods (which allow a deeper understanding of non--linear effects) are needed.  
Moreover, in scenarios 
radically different from $\Lambda\text{CDM}$\,, such as modified gravity or massive 
neutrinos, N--body simulations are still in their infancy. In particular, 
in the case of clustering dark energy with arbitrary
sound speed, as far as we know, no N--body simulations have been performed yet. Therefore, this work represents the first investigation of 
non--linearities in these scenarios. 

Our main finding is that the dependence of the 
non--linear correction to the power spectrum on the speed of sound is below 
the percent level in the baryon acoustic oscillation range of scales. This is good news for an efficient modelling of the power spectrum 
in applications such as parameter forecasting via, e.g. Markov chains. 
Indeed, once the effect of the sound speed is taken into account at the 
linear level, the non--linear part can be computed with good accuracy once and for all with the model of smooth dark energy with the same equation of state. We have also found that the effect of the non--linearities on the matter growth index must be taken into account for small scales for a reliable use of this parameter in future applications. 

The observational perspectives for these models are quite challenging but promising. The 
cleanest way to probe this type of scenarios is, in our opinion, 
through tomographic measurements of the gravitational potential via 
cosmic shear. Indeed, the discriminating power of galaxy surveys is 
probably limited, since the bias issue is further complicated by 
the emergence of a new clusterized component at late times. On the other 
hand, since lensing measurements typically probe the power spectrum at 
wavenumbers $k >1-10\, h\, \text{Mpc}^{-1}$\,, an accurate assessment of the non--linear 
effects is needed. This paper represents a first step in this 
direction.

\appendix

\section{A single fluid treatment for zero sound speed} \label{sv}
For the strict case of zero sound speed of dark energy, a treatment of the perturbations in terms of just two fields is proposed in \cite{Sefusatti:2011cm}\,. The corresponding equations are

\begin{align} 
\dot\delta_T(\mathbf{k})  &+ C(\tau)  \theta(\mathbf{k})+\int d^3\mathbf{p}\,d^3\mathbf{q}\,\delta_D(\mathbf{k}-\mathbf{p}-\mathbf{q})\alpha(\mathbf{q},\mathbf{p})\theta(\mathbf{q})\delta_T(\mathbf{p})=0 \label{ctot}\\
\dot\theta(\mathbf{k}) &+ \ch \theta(\mathbf{k}) + \frac{3}{2} \Omega_m \ch^2 \delta_T(\mathbf{k}) +\int d^3\mathbf{p}\,d^3\mathbf{q}\,\delta_D(\mathbf{k}-\mathbf{p}-\mathbf{q})\beta(\mathbf{q},\mathbf{p})\theta(\mathbf{q})\theta(\mathbf{p})=0\,, \label{etot}
\end{align}
where $\delta_T$ is related to the total clustering density perturbation $\delta_{tot}$
\be 
\delta_T \equiv \frac{\delta \rho}{\bar \rho_m} = \delta_m +  \delta_x \frac{\Omega_x}{\Omega_m}=\frac{\delta_{tot}}{\Omega_m} \label{dsv}
\ee
The variable $\theta$ is the common velocity perturbation of matter and dark energy (the same for both fluids)
and
\be\label{fc}
C(\tau) \equiv  1+ (1+w)\frac{\Omega_x}{\Omega_m}\,.
\ee
It is instructive to compare these equations with our non--linear formalism. Looking at the equation \eq{correuler} it is not obvious that for a zero sound speed of dark energy \eq{soundspeed} the velocity divergences of dark matter and dark energy are equal. However, this can be seen using the following expression for the rest frame sound speed in the Newtonian approximation:
\be  \label{soundn}
\nabla P +\vel \dot P=\cshat\left(\nabla \rho + \vel \dot \rho\right)\,,
\ee
which derives from the more general covariant definition \cite{Sefusatti:2011cm}\,:
\be \label{soundn2}
\left(g^{\mu\nu}+u^\mu u^\nu\right)\partial_\nu P=\cshat\left(g^{\mu\nu}+u^\mu u^\nu\right)\partial_\nu\rho
\ee
The equation \eq{soundn} can be obtained neglecting metric perturbations and combining the time and spatial components of \eq{soundn2}\,. It can be easily checked that \eq{soundn} gives consistently the transformation \eq{rf} when expanded at first order. Clearly, if $\cshat$ in \eq{soundn} is zero, the equation \eq{correuler} reduces to that of matter and therefore $\theta\equiv\theta_x=\theta_m$\,, provided that the initial conditions are the same for both quantities.

In reference \cite{Sefusatti:2011cm} the dark energy equation for zero sound speed is written as
\begin{align}
\label{eqcsvz}
\dot\delta(\mathbf{k})-3w\ch\delta(\mathbf{k})+(1+w)\theta(\mathbf{k})+\int  d^3\mathbf{p}\,d^3\mathbf{q}\,\delta_D(\mathbf{k}-\mathbf{p}-\mathbf{q})\alpha(\mathbf{q},\mathbf{p})\theta({q})\delta({p})=0\,,
\end{align}
which combined with its analogous for $w=0$ gives exactly \eq{ctot}\,. However, if we set $\cshat=0$ in \eq{eqc} we see that the result differs from \eq{eqcsvz} in a linear $\theta$--term and some $\mathcal{O}(2)$ terms which are not included in \eq{eqcsvz}\,. We will now explain these differences.
 
The continuity equation \eq{corrcont} can be rewritten as
\begin{align} \label{corrcont2}
\dot\rho+2\,\vel\cdot\dot\vel P+3\ch(\rho+P)+\nabla\cdot(\rho\vel)+P\,\nabla\cdot\vel+\vel\cdot\left(\nabla P+\vel \dot P\right)=0\,,
\end{align}
whose last term on the left hand side vanishes for zero rest frame sound speed. In the derivation of \eq{ctot} given in \cite{Sefusatti:2011cm}\,, the terms $2\vel\cdot\dot\vel P$ and $P \nabla\cdot\vel$ are both neglected, but it is clear that (at least) the second one of them cannot be discarded because it is of the same order as $\nabla\cdot(\rho\vel)$\,. Actually, making the correct sound speed assignment at first order with \eq{rf} and keeping the term $P \nabla\cdot\vel$\,, but neglecting $2\vel\cdot\dot\vel P$ (which gives subdominant contributions)\,, we obtain the equation \eq{eqcsvz} plus a linear correction in $\theta$ and a non--linear one in $\theta\times\theta$ which are both suppressed by $\mathcal{O}\left(\ch^2/k^2\right)$\,. On the other hand, neglecting the term $(2\vel\cdot\dot\vel P)$ in \eq{corrcont2}\,,  amounts at second order to take $-2 P\vel\left(\ch\vel+\nabla\phi_N\right)=0$\,, as it can be directly read from \eq{continuity} or \eq{correuler}. This approximation means setting to zero several second order non--linear contributions to the perturbed continuity equation that are proportional to $w$ (and hence vanish for dust but not for dark energy). With some algebra, one can check that these corrections are also suppressed at small scales by at least one power of $\ch^2/k^2$ (where $k$ refers to any non--linear combination of momenta) with respect to the usual non--linear $\alpha$ term\,. In conclusion, in order to get \eq{eqcsvz} and hence \eq{ctot} we just have to discard linear and non--linear terms in the perturbed continuity equation for dark energy that are small at scales $k\gg\ch$\,. Notice, incidentally, that to compute the precise form of these negligible second order corrections we would need to know \eq{rf} at second order; but, however, the linear one coming from $P \nabla\cdot\vel$ can be readily obtained (and it is a term $-9w(1+w)\ch^2\theta/k^2$ on the right hand side of \eq{eqcsvz})\,.

Notice that if we take \eq{2fmeul} and replace $\delta_x^L$ and $\delta_m^L$ by their non--linear counterparts, we obtain precisely \eq{etot}\,. The equations \eq{ctot} and \eq{etot} are a good approximation for the non--linear evolution of the total density perturbation in the case of dark energy with zero sound speed, as the results of Section \ref{numerical} show. This approximation is used in \cite{D'Amico:2011pf} to apply the TRG for the computation of the total power spectrum in the case $\cshat=0$ of maximal clustering \symbolfootnote[2]{While working in our project we learnt that a related study was being done simultaneously \cite{D'Amico:2011pf}\,.}.

\section*{Acknowledgments}
We thank Guido D'Amico and Emiliano Sefusatti for discussions. We also thank Toni Riotto, Sabino Matarrese, Filippo Vernizzi and Francis Bernardeau for various useful conversations and comments. G.B. thanks the Centro Enrico Fermi, and INFN for support. 

\bibliographystyle{hieeetr}
\bibliography{biblos}

\begin{thebibliography}{10}

\bibitem{Riess:1998cb}
A.~G. Riess {\em et~al.}, ``{Observational Evidence from Supernovae for an
  Accelerating Universe and a Cosmological Constant},'' {\em Astron. J.},
  vol.~116, pp.~1009--1038, 1998, astro-ph/9805201.

\bibitem{Perlmutter:1998np}
S.~Perlmutter {\em et~al.}, ``{Measurements of Omega and Lambda from 42
  High-Redshift Supernovae},'' {\em Astrophys. J.}, vol.~517, pp.~565--586,
  1999, astro-ph/9812133.

\bibitem{Komatsu:2010fb}
E.~Komatsu {\em et~al.}, ``{Seven-Year Wilkinson Microwave Anisotropy Probe
  (WMAP) Observations: Cosmological Interpretation},'' {\em Astrophys. J.
  Suppl.}, vol.~192, p.~18, 2011, 1001.4538.

\bibitem{Suzuki:2011hu}
N.~Suzuki {\em et~al.}, ``{The Hubble Space Telescope Cluster Supernova Survey:
  V. Improving the Dark Energy Constraints Above z>1 and Building an
  Early-Type-Hosted Supernova Sample},'' 2011, 1105.3470.

\bibitem{Sherwin:2011gv}
B.~D. Sherwin {\em et~al.}, ``{The Atacama Cosmology Telescope: Evidence for
  Dark Energy from the CMB Alone},'' 2011, 1105.0419.

\bibitem{Bonvin:2006en}
C.~Bonvin, R.~Durrer, and M.~Kunz, ``{The dipole of the luminosity distance: a
  direct measure of H(z)},'' {\em Phys. Rev. Lett.}, vol.~96, p.~191302, 2006,
  astro-ph/0603240.

\bibitem{Ballesteros:2008qk}
G.~Ballesteros and A.~Riotto, ``{Parameterizing the Effect of Dark Energy
  Perturbations on the Growth of Structures},'' {\em Phys. Lett.}, vol.~B668,
  pp.~171--176, 2008, 0807.3343.

\bibitem{Dave:2002mn}
R.~Dave, R.~Caldwell, and P.~J. Steinhardt, ``{Sensitivity of the cosmic
  microwave background anisotropy to initial conditions in quintessence
  cosmology},'' {\em Phys.Rev.}, vol.~D66, p.~023516, 2002, astro-ph/0206372.

\bibitem{Sandvik:2002jz}
H.~Sandvik, M.~Tegmark, M.~Zaldarriaga, and I.~Waga, ``{The end of unified dark
  matter?},'' {\em Phys.Rev.}, vol.~D69, p.~123524, 2004, astro-ph/0212114.

\bibitem{Bassett:2002fe}
B.~A. Bassett, M.~Kunz, D.~Parkinson, and C.~Ungarelli, ``{Condensate cosmology
  - Dark energy from dark matter},'' {\em Phys.Rev.}, vol.~D68, p.~043504,
  2003, astro-ph/0211303.

\bibitem{DeDeo:2003te}
S.~DeDeo, R.~R. Caldwell, and P.~J. Steinhardt, ``{Effects of the sound speed
  of quintessence on the microwave background and large scale structure},''
  {\em Phys. Rev.}, vol.~D67, p.~103509, 2003, astro-ph/0301284.

\bibitem{Doran:2003xq}
M.~Doran, C.~M. Muller, G.~Schafer, and C.~Wetterich, ``{Gauge-invariant
  initial conditions and early time perturbations in quintessence universes},''
  {\em Phys. Rev.}, vol.~D68, p.~063505, 2003, astro-ph/0304212.

\bibitem{Amendola:2003bz}
L.~Amendola, F.~Finelli, C.~Burigana, and D.~Carturan, ``{WMAP and the
  generalized Chaplygin gas},'' {\em JCAP}, vol.~0307, p.~005, 2003,
  astro-ph/0304325.

\bibitem{Afshordi:2006ad}
N.~Afshordi, D.~J.~H. Chung, and G.~Geshnizjani, ``{Cuscuton: A Causal Field
  Theory with an Infinite Speed of Sound},'' {\em Phys. Rev.}, vol.~D75,
  p.~083513, 2007, hep-th/0609150.

\bibitem{Kunz:2006wc}
M.~Kunz and D.~Sapone, ``{Crossing the Phantom Divide},'' {\em Phys.Rev.},
  vol.~D74, p.~123503, 2006, astro-ph/0609040.

\bibitem{Creminelli:2008wc}
P.~Creminelli, G.~D'Amico, J.~Norena, and F.~Vernizzi, ``{The Effective Theory
  of Quintessence: the $w < -1$ Side Unveiled},'' {\em JCAP}, vol.~0902,
  p.~018, 2009, 0811.0827.

\bibitem{Avelino:2008zz}
P.~Avelino, L.~Beca, and C.~Martins, ``{Linear and nonlinear instabilities in
  unified dark energy models},'' {\em Phys.Rev.}, vol.~D77, p.~063515, 2008.

\bibitem{Avelino:2008cu}
P.~Avelino, L.~Beca, and C.~Martins, ``{Clustering properties of dynamical dark
  energy models},'' {\em Phys.Rev.}, vol.~D77, p.~101302, 2008, 0802.0174.

\bibitem{Sapone:2009mb}
D.~Sapone, M.~Kunz, and M.~Kunz, ``{Fingerprinting Dark Energy},'' {\em Phys.
  Rev.}, vol.~D80, p.~083519, 2009, 0909.0007.

\bibitem{Kunz:2009yx}
M.~Kunz, A.~R. Liddle, D.~Parkinson, and C.~Gao, ``{Constraining the dark
  fluid},'' {\em Phys.Rev.}, vol.~D80, p.~083533, 2009, 0908.3197.

\bibitem{Lim:2010yk}
E.~A. Lim, I.~Sawicki, and A.~Vikman, ``{Dust of Dark Energy},'' {\em JCAP},
  vol.~1005, p.~012, 2010, 1003.5751.

\bibitem{Koshelev:2010wz}
N.~A. Koshelev, ``{Non-adiabatic perturbations in multi-component perfect
  fluids},'' {\em JCAP}, vol.~1104, p.~021, 2011, 1011.0569.

\bibitem{Novosyadlyj:2010pg}
B.~Novosyadlyj, O.~Sergijenko, S.~Apunevych, and V.~Pelykh, ``{Properties and
  uncertainties of scalar field models of dark energy with barotropic equation
  of state},'' {\em Phys. Rev.}, vol.~D82, p.~103008, 2010, 1008.1943.

\bibitem{Ansari:2011wv}
R.~U.~H. Ansari and S.~Unnikrishnan, ``{Perturbations in dark energy models
  with evolving speed of sound},'' 2011, 1104.4609.

\bibitem{Bean:2003fb}
R.~Bean and O.~Dore, ``{Probing dark energy perturbations: the dark energy
  equation of state and speed of sound as measured by WMAP},'' {\em Phys.
  Rev.}, vol.~D69, p.~083503, 2004, astro-ph/0307100.

\bibitem{Weller:2003hw}
J.~Weller and A.~M. Lewis, ``{Large Scale Cosmic Microwave Background
  Anisotropies and Dark Energy},'' {\em Mon. Not. Roy. Astron. Soc.}, vol.~346,
  pp.~987--993, 2003, astro-ph/0307104.

\bibitem{Hu:2004yd}
W.~Hu and R.~Scranton, ``{Measuring Dark Energy Clustering with CMB-Galaxy
  Correlations},'' {\em Phys. Rev.}, vol.~D70, p.~123002, 2004,
  astro-ph/0408456.

\bibitem{Hannestad:2005ak}
S.~Hannestad, ``{Constraints on the sound speed of dark energy},'' {\em Phys.
  Rev.}, vol.~D71, p.~103519, 2005, astro-ph/0504017.

\bibitem{Corasaniti:2005pq}
P.-S. Corasaniti, T.~Giannantonio, and A.~Melchiorri, ``{Constraining dark
  energy with cross-correlated CMB and Large Scale Structure data},'' {\em
  Phys. Rev.}, vol.~D71, p.~123521, 2005, astro-ph/0504115.

\bibitem{Kunz:2006ca}
M.~Kunz and D.~Sapone, ``{Dark energy versus modified gravity},'' {\em Phys.
  Rev. Lett.}, vol.~98, p.~121301, 2007, astro-ph/0612452.

\bibitem{Takada:2006xs}
M.~Takada, ``{Can A Galaxy Redshift Survey Measure Dark Energy Clustering?},''
  {\em Phys. Rev.}, vol.~D74, p.~043505, 2006, astro-ph/0606533.

\bibitem{Amendola:2007rr}
L.~Amendola, M.~Kunz, and D.~Sapone, ``{Measuring the dark side (with weak
  lensing)},'' {\em JCAP}, vol.~0804, p.~013, 2008, 0704.2421.

\bibitem{Mota:2007sz}
D.~F. Mota, J.~R. Kristiansen, T.~Koivisto, and N.~E. Groeneboom,
  ``{Constraining Dark Energy Anisotropic Stress},'' {\em Mon. Not. Roy.
  Astron. Soc.}, vol.~382, pp.~793--800, 2007, 0708.0830.

\bibitem{TorresRodriguez:2007mk}
A.~Torres-Rodriguez and C.~M. Cress, ``{Constraining the Nature of Dark Energy
  using the SKA},'' {\em Mon. Not. Roy. Astron. Soc.}, vol.~376,
  pp.~1831--1837, 2007, astro-ph/0702113.

\bibitem{Xia:2007km}
J.-Q. Xia, Y.-F. Cai, T.-T. Qiu, G.-B. Zhao, and X.~Zhang, ``{Constraints on
  the Sound Speed of Dynamical Dark Energy},'' {\em Int. J. Mod. Phys.},
  vol.~D17, pp.~1229--1243, 2008, astro-ph/0703202.

\bibitem{dePutter:2010vy}
R.~de~Putter, D.~Huterer, and E.~V. Linder, ``{Measuring the Speed of Dark:
  Detecting Dark Energy Perturbations},'' {\em Phys. Rev.}, vol.~D81,
  p.~103513, 2010, 1002.1311.

\bibitem{Ballesteros:2010ks}
G.~Ballesteros and J.~Lesgourgues, ``{Dark energy with non-adiabatic sound
  speed: initial conditions and detectability},'' {\em JCAP}, vol.~1010,
  p.~014, 2010, 1004.5509.

\bibitem{Sapone:2010uy}
D.~Sapone, M.~Kunz, and L.~Amendola, ``{Fingerprinting Dark Energy II: weak
  lensing and galaxy clustering tests},'' {\em Phys. Rev.}, vol.~D82,
  p.~103535, 2010, 1007.2188.

\bibitem{Li:2010ac}
H.~Li and J.-Q. Xia, ``{Constraints on Dark Energy Parameters from Correlations
  of CMB with LSS},'' {\em JCAP}, vol.~1004, p.~026, 2010, 1004.2774.

\bibitem{Ayaita:2011gp}
Y.~Ayaita, B.~M. Schaefer, and M.~Weber, ``{Investigating clustering dark
  energy with 3d weak cosmic shear},'' 2011, 1110.1985.

\bibitem{Bjaelde:2010qp}
O.~E. Bjaelde and Y.~Y.~Y. Wong, ``{Spherical collapse of dark energy with an
  arbitrary sound speed},'' 2010, 1009.0010.

\bibitem{Creminelli:2009mu}
P.~Creminelli, G.~D'Amico, J.~Norena, L.~Senatore, and F.~Vernizzi,
  ``{Spherical collapse in quintessence models with zero speed of sound},''
  {\em JCAP}, vol.~1003, p.~027, 2010, 0911.2701.

\bibitem{Sefusatti:2011cm}
E.~Sefusatti and F.~Vernizzi, ``{Cosmological structure formation with
  clustering quintessence},'' {\em JCAP}, vol.~1103, p.~047, 2011, 1101.1026.

\bibitem{Linder:2005in}
E.~V. Linder, ``{Cosmic growth history and expansion history},'' {\em Phys.
  Rev.}, vol.~D72, p.~043529, 2005, astro-ph/0507263.

\bibitem{Pietroni:2008jx}
M.~Pietroni, ``{Flowing with Time: a New Approach to Nonlinear Cosmological
  Perturbations},'' {\em JCAP}, vol.~0810, p.~036, 2008, 0806.0971.

\bibitem{Peebles:1980book}
P.~J.~E. Peebles, ``{The large--scale structure of the universe},'' {\em
  Princeton University Press}, 1980.

\bibitem{Wald:1984rg}
R.~M. Wald, ``{General Relativity},'' 1984.
\newblock Book, The University of Chicago Press, 1984.

\bibitem{Ma:1995ey}
C.-P. Ma and E.~Bertschinger, ``{Cosmological perturbation theory in the
  synchronous and conformal Newtonian gauges},'' {\em Astrophys. J.}, vol.~455,
  pp.~7--25, 1995, astro-ph/9506072.

\bibitem{Bernardeau:2001qr}
F.~Bernardeau, S.~Colombi, E.~Gaztanaga, and R.~Scoccimarro, ``{Large scale
  structure of the universe and cosmological perturbation theory},'' {\em
  Phys.Rept.}, vol.~367, pp.~1--248, 2002, astro-ph/0112551.

\bibitem{Lesgourgues:2009am}
J.~Lesgourgues, S.~Matarrese, M.~Pietroni, and A.~Riotto, ``{Non-linear Power
  Spectrum including Massive Neutrinos: the Time-RG Flow Approach},'' {\em
  JCAP}, vol.~0906, p.~017, 2009, 0901.4550.

\bibitem{Anselmi:2010fs}
S.~Anselmi, S.~Matarrese, and M.~Pietroni, ``{Next-to-leading resummations in
  cosmological perturbation theory},'' {\em JCAP}, vol.~1106, p.~015, 2011,
  1011.4477.

\bibitem{Bartolo:2009rb}
N.~Bartolo, J.~P.~B. Almeida, S.~Matarrese, M.~Pietroni, and A.~Riotto,
  ``{Signatures of Primordial non-Gaussianities in the Matter Power-Spectrum
  and Bispectrum: the Time-RG Approach},'' {\em JCAP}, vol.~1003, p.~011, 2010,
  0912.4276.

\bibitem{Brouzakis:2010md}
N.~Brouzakis, V.~Pettorino, N.~Tetradis, and C.~Wetterich, ``{Nonlinear matter
  spectra in growing neutrino quintessence},'' {\em JCAP}, vol.~1103, p.~049,
  2011, 1012.5255.

\bibitem{Brouzakis:2010hp}
N.~Brouzakis and N.~Tetradis, ``{Non-linear Matter Spectrum for a Variable
  Equation of State},'' {\em JCAP}, vol.~1101, p.~024, 2011, 1002.3277.

\bibitem{Saracco:2009df}
F.~Saracco, M.~Pietroni, N.~Tetradis, V.~Pettorino, and G.~Robbers,
  ``{Non-linear Matter Spectra in Coupled Quintessence},'' {\em Phys.Rev.},
  vol.~D82, p.~023528, 2010, 0911.5396.

\bibitem{Elia:2010en}
A.~Elia, S.~Kulkarni, C.~Porciani, M.~Pietroni, and S.~Matarrese, ``{Modelling
  the clustering of dark matter haloes in resummed perturbation theories},''
  {\em Monthly Notices of the Royal Astronomical Society}, vol.~416, no.~3,
  pp.~1703--1716, 2011, 1012.4833.

\bibitem{Lewis:1999bs}
A.~Lewis, A.~Challinor, and A.~Lasenby, ``{Efficient computation of CMB
  anisotropies in closed FRW models},'' {\em Astrophys.J.}, vol.~538,
  pp.~473--476, 2000, astro-ph/9911177.

\bibitem{Wang:1998gt}
L.-M. Wang and P.~J. Steinhardt, ``{Cluster Abundance Constraints on
  Quintessence Models},'' {\em Astrophys. J.}, vol.~508, pp.~483--490, 1998,
  astro-ph/9804015.

\bibitem{D'Amico:2011pf}
G.~D'Amico and E.~Sefusatti, ``{The nonlinear power spectrum in clustering
  quintessence cosmologies},'' 2011, 1106.0314.

\end{thebibliography}

\end{document}